\documentclass[10pt]{article}
\usepackage[nosort]{cite}
\usepackage{geometry}
\usepackage{graphicx}
\usepackage{slashed}
\usepackage{amsmath}
\usepackage{verbatim}
\usepackage{url}
\usepackage{amsfonts}
\usepackage{layout}
\usepackage{slashed}
\usepackage{dcolumn}
\usepackage{bm}
\usepackage{multirow}

\usepackage{amssymb}
\newcommand{\ldotss}{\!.\,.\,}

\newcommand{\myperp}{{{\perp}}}

\newcommand{\myhline}{\noalign{\smallskip}\hline\noalign{\smallskip}}

\newcommand{\barW}{{\overline{W\!}\,}}

\renewcommand{\baselinestretch}{1.05}
\newcommand{\dash}{\text{-}}

\newcommand{\sym}{{\rm sym}}

\newcommand{\ie}{{\it i.e.}\ }

\newcommand{\cn}{{\cal N}}
\newcommand{\be}{\begin{equation}}
\newcommand{\bea}{\begin{eqnarray}}
\newcommand{\beq}{\begin{equation}}

\newcommand{\delete}[1]{}

\newcommand{\ee}{\end{equation}}
\newcommand{\eea}{\end{eqnarray}}
\newcommand{\eeq}{\end{equation}}
\newcommand{\lsim}{\!\mathrel{\hbox{\rlap{\lower.55ex \hbox{$\sim$}} \kern-.34em \raise.4ex \hbox{$<$}}}}
\newcommand{\gsim}{\!\mathrel{\hbox{\rlap{\lower.55ex \hbox{$\sim$}} \kern-.34em \raise.4ex \hbox{$>$}}}}

\newcommand{\<}{\langle}
\renewcommand{\>}{\rangle}

\begin{document}

\begin{titlepage}

\begin{flushright}
 PUPT-2378\\
 NSF-KITP-11-094
\end{flushright}
\vspace{4cm}

\begin{center}
{\Large \bf The Coulomb-branch S-matrix from massless amplitudes}

\vspace{1cm}
{\bf  Michael Kiermaier}\\
\vspace{1cm}
{\it Department of Physics, Princeton University \\ Princeton, NJ 08544, USA}\\[1.3ex]
and\\[1.3ex]
{\it Kavli Institute for Theoretical Physics, University of California\\
Santa Barbara, CA 93106, USA}
\end{center}
\vspace{1.5cm}

\begin{abstract}
We present a systematic method to extract the entire tree-level S-matrix on the Coulomb branch of $\,\cn=4$ SYM from soft-scalar limits of on-shell amplitudes at the origin of moduli space. Massive amplitudes in the spontaneously-broken theory can thus be computed from on-shell amplitudes in the massless, unbroken theory.
To check this correspondence, we first prove  that soft and collinear divergences in the required massless amplitudes cancel for a judicious choice of soft-scalar momenta. We then explicitly verify  our proposal in examples with arbitrarily many external legs and to all orders in the mass. As a byproduct, the construction leads to a massive CSW-like expansion  that reproduces several known all-$n$ results for Coulomb-branch amplitudes in an effortless way. We briefly discuss the extension of our method to loop integrands.

\end{abstract}

\end{titlepage}

\tableofcontents

\newpage

\setcounter{equation}{0}
\section{Introduction}
The study of field theories with massless particles has led to remarkably compact expressions for on-shell amplitudes at both tree- and loop level, most notably in planar $\cn\!=\!4$ super Yang-Mills theory at the origin of moduli space~\cite{Parke:1986gb,Drummond:2008vq,Brandhuber:2008pf,Drummond:2008bq,Drummond:2008cr,Elvang:2008vz,Kiermaier:2009yu,Brandhuber:2009xz,Elvang:2009ya,ArkaniHamed:2009vw,ArkaniHamed:2010kv,Bullimore:2010dz,Mason:2010yk}.
It is natural to ask whether results for massless on-shell amplitudes  can be ``recycled'' to compute scattering processes with {\em massive} particles. 
Massive amplitudes are of interest for several reasons. Not only are massive particles ubiquitous in phenomenologically relevant theories;  masses are also used in massless theories to regulate infrared divergences, and to facilitate the computation of rational terms by encoding the $D>4$ components of loop momenta in $D$-dimensional unitarity cuts~\cite{Badger:2008cm}.\footnote{This is used in numerical implementations of the unitarity method~\cite{Bern:1994zx,Bern:1994cg}, such as \textsc{BlackHat}~\cite{Berger:2008sj}.}

In this paper, we argue that massless planar on-shell amplitudes at the origin of moduli space of $\cn\!=\!4$ SYM can be used to systematically compute the tree-level S-matrix of $\cn\!=\!4$ SYM  on the Coulomb branch.\footnote{Coulomb-branch amplitudes have recently also been studied in~\cite{Schabinger:2008ah, Boels:2010mj,Craig:2011ws,Boels:2011zz}. In~\cite{Alday:2009zm, Henn:2010bk,Henn:2010ir,Henn:2011xk}, they were used to regularize IR divergences in massless amplitudes (see also~\cite{Mitov:2006xs,Sever:2009aa}).} Massive amplitudes in the spontaneously-broken theory are thus determined by on-shell amplitudes in the massless, unbroken theory.
 We also propose that this relation extends to the loop integrand in a certain large-$N$ limit. 

As a simple example, consider the Coulomb-branch scattering of two adjacent massive `W-bosons' of mass $m^2$,  with an arbitrary number of massless particles.
As the Coulomb-branch is parameterized by vacuum expectation values $\<\phi_{ab}\>$ of scalar fields, it is natural to expect that the Coulomb-branch amplitude is related to soft-scalar limits of massless color-ordered amplitudes at the origin of moduli space~\cite{Craig:2011ws}. Therefore, we might naively guess
\begin{equation}
\begin{split}\label{guess}
    \bigl\<\,W_1\barW_2\,\ldots\,\bigr\>~~\overset{?}{=}~~\lim_{\varepsilon\to 0}\,\sum_{s=0}^\infty\, \bigl\<\,g_1\,\phi^{\rm vev}_{\varepsilon q_1}\phi^{\rm vev}_{\varepsilon q_2}\,.\,.\,\phi^{\rm vev}_{\varepsilon q_s}\,g_2\,\ldots\,\bigr\>\,,
\end{split}
\end{equation}
where  $W,\barW$ and $g$ are W-bosons and massless gluons, respectively, while  $\phi^{\rm vev}_{\varepsilon q_i}$ is a massless scalar\footnote{We denote such soft scalars with R-symmetry indices in the vev direction as ``vev scalars'' in the following. From the point of view of the massless theory, which sits at the origin of moduli space and does not `know' about the vev, these are regular scalar external states of the on-shell amplitude.} of momentum $\varepsilon q_i$ whose R-symmetry structure is oriented in the vev direction, $\phi^{\rm vev}=\<\phi_{ab}\>\phi^{ab}$. The `$\ldots$' in~(\ref{guess}) denote arbitrary additional massless states in the amplitude. We inserted scalars $\phi^{\rm vev}$ between $g_1$ and $g_2$ on the right-hand side of~(\ref{guess}), because this is the only allowed position considering the color-trace structure of the Coulomb-branch amplitude~\cite{Craig:2011ws}.
As the scalar vev determines the mass of W-bosons,  $m\sim \<\phi\>$, the proposal~(\ref{guess}) can be interpreted as a mass expansion of the Coulomb-branch amplitude. However,~(\ref{guess}) cannot be quite right as it stands; at the very least, additional information must be supplied to make it well-defined, as the following puzzles illustrate: (1)~The on-shell momenta on lines $1$ and $2$ are massive on the left-hand side, but they are massless on the right-hand side. How are these momenta related to each other? \,(2)~While gluon helicities are Lorentz invariant, the polarizations of massive vector bosons depend on the frame. How does this frame-dependence show up on the right-hand side? \,(3)~The soft limit $\varepsilon\to0$ on the right-hand side  depends on the momentum directions $q_i$ of the vev scalars; how does this $q_i$-dependence reflect itself in the Coulomb-branch amplitude? (4)~Finally, there is another, even more worrisome obstacle: the soft-scalar amplitudes on the right-hand side generically suffer from soft divergences in the limit $\varepsilon\to0$, which get worse as we increase the number of soft scalars $s$. How can these divergences be avoided or treated?

In~\cite{Craig:2011ws}, the first three puzzles above were addressed and resolved as follows. One first introduces an arbitrary light-like reference vector $q$. Motivated by the `massive spinor-helicity formalism'~\cite{Kleiss:1985yh,Dittmaier:1998nn}\footnote{We use the conventions summarized in appendix B of~\cite{Cohen:2010mi}.}, we decompose the momenta on the massive lines of the Coulomb-branch amplitude as
\begin{equation}\label{decomp}
p_i~=~p_i^\myperp-\frac{m_i^2}{2\,q\!\cdot\! p_i}q\,.
\end{equation}
In this decomposition, $p_i^\myperp$ is a light-like momentum that differs from the massive momentum $p_i$ only by an $O(m^2)$ term. It is thus natural to pick $p_i^\perp$ as the massless momentum of the gluon $g_i$ on the right-hand side of~(\ref{guess}). We also use $q$ as the reference vector for the $W$-boson polarizations. The frame-dependence of the Coulomb-branch amplitude then reflects itself in an explicit $q$-dependence on the left-hand side of~(\ref{guess}). We have thus specified how momenta and polarizations of massive and massless particles on both sides of~(\ref{guess}) relate to each other; for a more detailed dictionary, including the explicit polarization vectors, the interested reader can peek ahead at Table~\ref{tabstates} below. The left-hand side of~(\ref{guess}) now depends on the choice of one light-like vector $q$. For~(\ref{guess}) to pass the basic consistency check of relating objects with the same parameters, we are forced to pick all vev-scalar momenta on the right-hand side to be identical, $q_i\to q$. But this introduces a new problem: the right-hand side  suffers from collinear divergences in this limit! In~\cite{Craig:2011ws}, a remedy to this problem was proposed: since collinear divergences are anti-symmetric in collinear momenta, we need to symmetrize the right-hand side of~(\ref{guess}) in all vev-scalar momenta $q_i$ before taking the limit $q_i\to q$.

In summary, this suggests a refined version of the proposal~(\ref{guess}):
\begin{equation}\label{generalization}
    \boxed{\phantom{\Biggl(}\!
    \bigl\<\,W_1\barW_2\,\ldots\,\bigr\>~~=~~\lim_{\varepsilon\to 0}\,\sum_{s=0}^\infty\,\, \bigl\<\,g_1\,\underbrace{\phi^{\rm vev}_{\varepsilon q}\ldotss\phi^{\rm vev}_{\varepsilon q}}_{s
    \text{ times}}\,g_2\,\ldots\,\bigr\>_{\rm sym}\,.~~}
\end{equation}
Here, the subscript `sym' denotes a symmetrization of the scalars in their momentum directions $q_i$ before taking the collinear limit $q_i\to q$. In the Coulomb-branch amplitude, the massive states have momenta $p_i$  and $q$-dependent polarizations; they are mapped to massless particles with momenta $p_i^\perp$.
It was argued in~\cite{Craig:2011ws} that~(\ref{generalization}) can be used to compute the Coulomb-branch amplitude {\em to leading order in the mass}. Beyond the leading non-vanishing term in the sum over $s$, however, (\ref{generalization}) generically suffers from soft divergences in the limit $\varepsilon\to 0$. We will argue in this paper that {\em all soft divergences cancel for a judicious choice of the scalar momentum direction $q$.} For this choice of $q$,~(\ref{generalization}) is valid to all orders, as we verify in several examples. We then generalize~(\ref{generalization}) to amplitudes with arbitrarily many massive lines of arbitrary mass, and to the Coulomb-branch loop integrand in a certain large-$N$ limit.

One can actually deduce the special choice of $q$ for which~(\ref{generalization}) holds from another point of view, without ever considering soft divergences. In fact, there is something puzzling about the proposal~(\ref{generalization}) as we have stated it so far: the momenta $p_i^\perp$ of the particles in the massless amplitude on the right-hand side do not sum to zero! In fact, it follows directly from~(\ref{decomp}) that momentum conservation is violated on the right-hand side by a term of $O(m^2)$. Clearly, this violation is subleading in the mass expansion, and was thus irrelevant for the leading-order analysis of~\cite{Craig:2011ws}. To get the correct Coulomb-branch amplitude to all orders from~(\ref{generalization}), however, we cannot neglect this subtlety: the subleading terms are ambiguous, and~(\ref{generalization}) is thus generically ill-defined. There is a surprisingly simple cure for this problem: we simply demand that $q$ is chosen in such a way that the massless projections $p_i^\perp$ of the momenta in the Coulomb-branch amplitude sum to zero.\footnote{This choice of $q$ was used in~\cite{Craig:2011ws} to simplify expressions for Coulomb-branch superamplitudes.} Explicitly, we demand that $q$ satisfies
\begin{equation}\label{spq}
    \sum_{i=1}^n\frac{m_i^2}{2\,q\!\cdot\! p_i}~=~0\qquad\qquad\qquad\text{ (special choice of $q$)}\,.
\end{equation}
For Coulomb-branch amplitudes with two adjacent massive lines, which satisfy $m_1^2=m_2^2$ due to the Coulomb-branch constraint $\sum_i m_i=0$, this special choice of $q$ is equivalent to the simple orthogonality condition $q\!\cdot\! (p_1\!+\!p_2)=0$ (while keeping $q\!\cdot\! p_1=-q\!\cdot\!p_2\neq 0$). The condition becomes more complicated as we increase the number of massive lines, but a solution always\footnote{Only for $n=3$ is~(\ref{spq}) problematic because, for example, $q\!\cdot\! p_1=-q\!\cdot\!p_2$ implies $q\!\cdot\!p_3=0$ so that $q$ is a singular choice of reference vector for line $3$.}  exists. In practice, we will not need to solve for a $q$ that satisfies~(\ref{spq}); instead,~(\ref{spq}) simply provides an additional relation that can be used to manipulate amplitudes.

Remarkably, with the choice~(\ref{spq}) for $q$, {\em all soft divergences cancel} in the symmetrized  soft-scalar amplitudes
$\<g_1\phi^{\rm vev}_{\varepsilon q}\ldotss\phi^{\rm vev}_{\varepsilon q}g_2\ldots\>_{\rm sym}$\,, for any number of vev scalars!
The proof of the cancellation of soft divergences in~(\ref{generalization}), and in the corresponding generalization to arbitrary Coulomb-branch amplitudes, is one of the main results of this paper. This makes the proposal well-defined to all orders, and allows us to verify it in examples.

As we stated above, the symmetrization in~(\ref{generalization}) is intended to cancel the collinear divergences in the soft-scalar amplitudes. Indeed, we will show explicitly below that all collinear divergences  cancel after symmetrization in the momenta $\varepsilon q_i$ of the vev scalars.  This symmetrization is actually quite natural, independent of its purpose to eliminate collinear divergences: as the color generators of scalars in the vev direction commute with each other, vev scalars should not be ordered with respect to one another. This ``unorderedness'' is accomplished by the sum over permutations in the color-ordered amplitude. However, one may fear that the  summation over $s!$ terms in the symmetrization of an amplitude with $s$ vev scalars is so cumbersome that it makes the proposal~(\ref{generalization}) impractical to implement. As we will see, the exact opposite is the case: the {\em sum over symmetrizations simplifies the amplitude} for a convenient choice of representation. Indeed, choosing the MHV vertex expansion~\cite{Cachazo:2004by,Risager:2005vk,Elvang:2008na,Elvang:2008vz,Bullimore:2010dz}  with reference spinor $|q]$, we will show that the symmetrization eliminates many diagrams from the expansion, and simplifies others.

To obtain the full Coulomb-branch amplitude from the proposal~(\ref{generalization}), the subleading terms on the right-hand side in~(\ref{generalization}) in particular must reproduce the subleading terms in internal propagators $1/(P_I^2+m^2)$. In field theory, it is well-known that the `$+m^2$' in a propagator can be built up as a sum over diagrams in which we treat the mass term in the Lagrangian as an interaction vertex. In our case, however, there is a key difference: $P_{I}^2$ itself also contains an $O(m^2)$ piece because $P_{I}=\sum (p_i^\perp\!-\!\frac{m_i^2}{2q\cdot p_i}q)$. This is more than a technicality; as we are only allowing ourselves to use {\em on-shell} amplitudes of the massless theory in the construction of the Coulomb-branch S-matrix, both external and internal momenta  differ between the massless and the massive scattering process. If the proposal~(\ref{generalization}) is correct beyond the leading order, it must account both for `$+m^2$' {\em and} the  shift of $P_{I}^2$ itself in the propagator. As we will see, the `$+m^2$' is indeed produced by attaching scalars directly to the internal propagator, while the $O(m^2)$ piece in $P_I^2$ is reproduced in a much more interesting way: it comes from the finite term that remains after summing all soft-divergent diagrams in which soft scalars attach to the external lines!

In this paper, we use the MHV vertex expansion as a technical tool to analyze the soft-scalar amplitudes on the right-hand side of~(\ref{generalization}).
It allows us to `resum' the infinite sum over soft-scalar amplitudes into expressions with a finite number of diagrams. The `Feynman rules' for these diagrams are easily deduced, leading to a new CSW-like expansion for Coulomb-branch amplitudes. CSW expansions for amplitudes in spontaneously-broken gauge theories were considered previously: in~\cite{Buchta:2010qr}, a  CSW expansion for the electro-weak theory was derived from a Lagrangian~\cite{Gorsky:2005sf,Mansfield:2005yd,Mason:2005kn} approach (see also~\cite{Boels:2007pj}). It is quite curious that our construction of a CSW-like expansion here neither relies on Lagrangians nor on complex shifts, but instead follows directly from~(\ref{generalization}). Despite the focus on CSW expansions in this work, we emphasize that~(\ref{generalization}) can be applied to any choice of representation for the soft-scalar amplitudes. Other choices may lead to new interesting expressions for Coulomb-branch amplitudes that make, for example, dual conformal symmetry manifest.

\medskip
{\em This paper is organized as follows.} In section~\ref{sec2mass}, we verify the proposal~(\ref{generalization}) in examples of Coulomb-branch amplitudes with two massive lines. In section~\ref{CSW2}, we derive the diagrammatic rules of the CSW-like expansion for two-mass amplitudes, and use it reproduce various known Coulomb-branch amplitudes. In section~\ref{secgeneral}, we present our proposal for Coulomb-branch amplitudes with arbitrary masses, and prove that all soft divergences cancel in general. The CSW-like expansion for general Coulomb-branch amplitudes is presented in section~\ref{CSWgen}. We take a peek at the loop-level generalization of our results in section~\ref{secloop}. We end with a discussion of promising future directions in section~\ref{secdisc}.

\setcounter{equation}{0}
\section{Amplitudes with two adjacent massive lines}\label{sec2mass}
In this section we illustrate the proposal~(\ref{generalization}) for Coulomb-branch amplitudes with 2 adjacent massive lines. As it turns out, these examples are rich enough to illustrate all the essential mechanisms that justify~(\ref{generalization}).
\subsection{Setup}
In this section, we consider $\cn\!=\!4$ SYM with gauge group $U(M\!+\!N)$, and we give vevs to a subset of the scalars,
\begin{equation}
    \big\<(\phi_{12})_A{}^{B}\big\>\, = \,\big\<(\phi_{34})_A{}^{B}\big\>  \,=\, m \, \delta_{A}{}^{B}\qquad \qquad\text{ for }~~~ 1\leq A,B\leq M\,.
\end{equation}
This breaks the gauge group spontaneously to $U(M)\!\times\! U(N)$, and the $R$-symmetry group as $SU(4)\to Sp(4)$. The massless sector --- namely fields $Y_A{}^B$ with $A,B$ both in $U(M)$ or both in $U(N)$ --- contains the familiar gluons $g^\pm$, fermions $\chi^a$, and scalars $\phi^{ab}$, where $a,b$ are R-symmetry indices. The massive sector, with fields $X_A{}^B$ that are bifundamental with respect to $U(M)\!\times\! U(N)$, consists of $W$ bosons, scalars $w$, and fermions $\psi$. Table~\ref{tabstates} summarizes the massless and massive states, their polarizations, and how they correspond to each other.
\begin{table}[t!]
\begin{center}
\begin{tabular}{r ccc }
&massless fields $Y$ & massive fields $X$ & massive polarisation/wavefct\\
\noalign{\smallskip}
\cline{2-4}\\[-2.5ex]
\cline{2-4}
\noalign{\medskip}
gluons / $W^\pm$-boson: & $g^+$, $g^-$ &  $W^+$, $W^-$
&$\epsilon^-\!=\!\frac{\sqrt{2}|i^\myperp\> [q|}{[ i^\myperp q]},~~
    \epsilon^+\!=\!\frac{\sqrt{2}|q\> [i^\myperp|}{\< i^\myperp q\>}$
\\
\myhline
scalar / $W^L$-boson: & $\tfrac{1}{\sqrt{2}}(\phi^{12}\!+\!\phi^{34})$ &  ~$W^L\!\sim\! \tfrac{1}{\sqrt{2}}(w^{12}\!+\!w^{34})$~
&
$\slashed{\epsilon}^L =
   \frac{1}{m_i} \Big( \slashed{p}_i^\myperp  + \frac{m_i^2}{2q\cdot p_i} \slashed{q} \Big)$
\\
\myhline
\multirow{2}{*}{scalars:} & ~~$\phi^{13}$, $\phi^{14}$, $\phi^{23}$, $\phi^{24}$,~~ & $w^{13}$, $w^{14}$, $w^{23}$, $w^{24}$, &\multirow{2}{*}{1}\\[.3ex]
&$\tfrac{1}{\sqrt{2}}(\phi^{12}\!-\!\phi^{34})$&$\tfrac{1}{\sqrt{2}}(w^{12}\!-\!w^{34})$&\\
\myhline
fermions:& $\chi^a$ & $\psi^a$ &
$\lambda^+=|i^\perp]\,,~~\tilde{\lambda}^-=|i^\perp\> $
\end{tabular}
\end{center}
\caption{\small Massless and massive particles on the Coulomb branch for the R-symmetry breaking~$SU(4)\!\to\!Sp(4)$.}
\label{tabstates}
\end{table}

As a warm-up, consider the  Coulomb-branch amplitude~\cite{Craig:2011ws}
\begin{equation}\label{4ptRviol}
    \bigl\< W^-_1 \barW^+_2\, \phi^{34}_3\, \phi^{34}_4 \bigr\>
   ~=~-\frac{m^{2}\<1^\perp|q|2^\perp]}{\<2^\perp|q|1^\perp](P_{23}^2+m^2)}\qquad\text{with }~P_{23}=p_2^\perp-\frac{m^2}{2\,q\!\cdot\! p_2}q +p_3\,.
\end{equation}
Here, massive external momenta are decomposed as in~(\ref{decomp}), and the spinors $|i^\perp]$ and $|i^\perp\>$ are associated with the massless momenta $p_i^\perp\!=\!|i^\perp\>[i^\perp|$. The polarization vectors of W-boson are defined with respect to the reference spinor $q$ as shown in Table~\ref{tabstates}.
Unlike in the massless case, massive amplitudes explicitly depend on the reference vector $q$ that appears in the polarization vectors.
Note that the  massless scalars $\phi^{34}$ in~(\ref{4ptRviol}) are regular external states of the Coulomb-branch amplitude, and are {\em not} the soft scalars that we will need to add to reconstruct this amplitude from data at the origin of moduli space. The amplitude~(\ref{4ptRviol}) is $O(m^2)$, and thus vanishes in the massless limit, because it violates the $SU(4)_R$ symmetry that is unbroken in the massless theory.

It is convenient to decompose internal momenta $P_I$, such as the $P_{23}$ that appears in the propagator of~(\ref{4ptRviol}), as
\begin{equation}
    P_I~=~P_{I}^\perp-\sum_{i\in I}\frac{m_i^2}{2\,q\!\cdot\! p_i}q\,, \qquad\quad \text{with }~~P_I^\perp\equiv\sum_{i\in I} p_i^\perp\,.
\end{equation}
This decomposes $P_I$ into an $O(1)$ piece (namely $P_I^\perp$) and an $O(m^2)$ piece.
Note that, with this definition, $P_I^\perp$ itself is not null; rather, it is the sum over the null projections $p_i^\perp$ associated with the external momenta $i\in I$.
We can now  carry out  a mass expansion of the amplitude~(\ref{4ptRviol}):
\begin{equation}\label{4ptexpand}
    \bigl\< W^-_1 \barW^+_2\, \phi^{34}_3\, \phi^{34}_4 \bigr\>
   ~=~
    -\frac{m^{2}\<1^\perp|q|2^\perp]}{\<2^\perp|q|1^\perp](P^\perp_{23})^2}\,\sum_{v=0}^\infty\Bigl(\frac{m^2\,q\!\cdot\! p_3}{(P^\perp_{23})^2\,q\!\cdot\! p_2}\Bigr)^v\,.
\end{equation}
This mass expansion is unambiguous for the ``special choice of $q$'',~(\ref{spq}), which for the two-mass case at hand amounts to demanding $q\!\cdot\!(p_1\!+\!p_2)\!=\!0$.

The all-$n$ generalization of the amplitude~(\ref{4ptRviol}) is~\cite{Craig:2011ws}
\begin{equation}\label{nptRviol}
    \bigl\< W^-_1 \barW^+_2\, \phi^{34}_3\ldots \phi^{34}_n \bigr\>
    ~=~\frac{-m^{n-2}\<1^\perp|q|2^\perp]}{\<2^\perp|q|1^\perp](P_{23}^2+m^2)(P_{234}^2+m^2)\cdots(P_{23\ldots n\dash1}^2+m^2)}\,.
\end{equation}
These ``maximally $SU(4)$-violating amplitudes'' are extremely simple, and therefore the ideal testing ground to check the proposal~(\ref{generalization}). For these amplitudes, the proposal reads
\begin{equation}\label{propn}
    \bigl\< W^-_1 \barW^+_2\, \phi^{34}_3\ldots \phi^{34}_n \bigr\>~~=~~\lim_{\varepsilon\to 0}\,\sum_{s=n-2}^\infty \,
    \bigl\<\,g_1^-\,\underbrace{\phi^{\rm vev}_{\varepsilon q}\ldotss\phi^{\rm vev}_{\varepsilon q}}_{s \text{ times}}\,g_2^+\,\phi^{34}_3\ldots\phi^{34}_n\,\bigr\>_{\rm sym}\,.
\end{equation}
The $W$ and $\barW$ vector bosons are bifundamentals and anti-bifundamentals of $U(M)\!\times\! U(N)$, respectively; the vev scalars $\phi^{\rm vev}$, which `live' in $U(M)$, are therefore only inserted between $g_1$ and $g_2$ on the right-hand side of~(\ref{propn}).
Explicitly, the vev scalars are given by
\begin{equation}\label{phivev}
    \phi^{\rm vev}~=~\<\phi_{ab}\>\phi^{ab}~=~m\bigl(\phi^{12}+\phi^{34}\bigr)\,.
\end{equation}
In~(\ref{propn}) we dropped massless amplitudes with $s<n\!-\!2$ vev scalars because these amplitudes violate $SU(4)_R$ and therefore vanish. The $s=n\!-\!2$ term is the first amplitude for which the vev scalars can `soak up' the $SU(4)_R$ violation introduced by the scalars $\phi^{34}$, and it was shown in~\cite{Craig:2011ws} that~(\ref{propn}) reproduces this leading-order term correctly.
As we will show in this section, it is possible to verify~(\ref{propn}) to all orders.

\subsection{Simplification through vev-scalar symmetrization}\label{seccoll2}
To verify~(\ref{propn}), we need to compute the symmetrized massless amplitudes
\begin{equation}\label{26}
    \bigl\<\,g_1\,\phi^{\rm vev}_{\varepsilon q}\ldotss\phi^{\rm vev}_{\varepsilon q}\,g_2\,\phi^{34}_3\ldots\phi^{34}_n\,\bigr\>_{\rm sym}
\end{equation}
 for any number of vev scalars $s$. The symmetrization is carried out by taking the $s$ vev scalars to have distinct momenta $\varepsilon q_i$, then averaging over the $s!$ permutations of these momenta. Finally, we then take the collinear limit $q_i\to q$. Note that throughout this part of the analysis, we do {\em not} yet take the vev scalars to be soft; $\varepsilon$ is not infinitesimal at this stage.

At first sight, summing over all symmetrizations seems a daunting task. It is tremendously simplified by expressing the required massless amplitudes using the MHV vertex expansion~\cite{Cachazo:2004by} with reference spinor $|q]$. The MHV vertex expansion computes amplitudes as a sum over diagrams with MHV amplitudes as vertices, connected by scalar propagators. For internal (and thus off-shell) lines $P_I$, CSW instruct us to compute MHV vertices using the spinor
\begin{equation}\label{CSW}
    |P_I\>~\equiv~P_I|q]\,.
\end{equation}
In the MHV vertex representation of the amplitude~(\ref{26}), the symmetrization in soft-scalar momenta only needs to be carried out MHV vertex by MHV vertex. Indeed, as we will see momentarily, symmetrizing
individual MHV vertices ensures that each diagram is either finite or vanishes; therefore, further symmetrizations of vev scalars between different MHV vertices will not affect the collinear limit.

As a simple example, consider the three-point MHV vertex with two vev scalars of momentum $\varepsilon q_1$ and $\varepsilon q_2$. We find
\begin{equation}\label{3ptsym}
\parbox[c]{1.5cm}{\includegraphics[width=1.3cm]{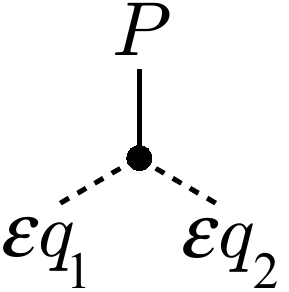}}_\sym~ ~\propto~~\frac{1}{\<Pq_1\>\<q_1q_2\>\<q_2 P\>}+\frac{1}{\<Pq_2\>\<q_2q_1\>\<q_1 P\>}~=~0\,.
\end{equation}
Here and in the following, we denote MHV vertices by solid dots, vev-scalar external states by dashed lines, and  internal lines or other non-vev-scalar external states by solid lines. Though each vev scalar in~(\ref{3ptsym}) is a linear combination of the form~(\ref{phivev}), the only non-vanishing contribution arises when one of the vev scalars is $\phi^{12}$ and the other one $\phi^{34}$. In~(\ref{3ptsym}), we dropped  the MHV numerator $\<...\>^4$, which is an overall factor because it only depends on the choice of particle on the third line, and not on the ordering of particles.  Importantly, the right-hand side of~(\ref{3ptsym}) vanishes identically, {\em even before taking the collinear limit $q_i\to q$.}
We  thus do not have to worry that the collinear divergence in the propagator $1/(q_1\!+\!q_2)^2$ that connects this vertex to the rest of the diagram might cancel this zero, resulting in a finite contribution.  It thus follows from~(\ref{3ptsym}) that we can drop all MHV vertex diagram that contain three-point vertices with two vev scalars.

Encouraged by this somewhat trivial example, we can compute the symmetrizations of all MHV vertices in which all lines except for one are vev scalars. As MHV vertices contain at most 4 scalars, such vertices exist for $n\leq5$. Explicit computation shows that all of these vanish after symmetrization, even before taking the collinear limit:
\begin{equation}\label{allptsym}
    \parbox[c]{1.5cm}{\includegraphics[width=1.3cm]{3ptsym}}_\sym=~0\,,\quad\qquad
    \parbox[c]{1.6cm}{\includegraphics[width=1.7cm]{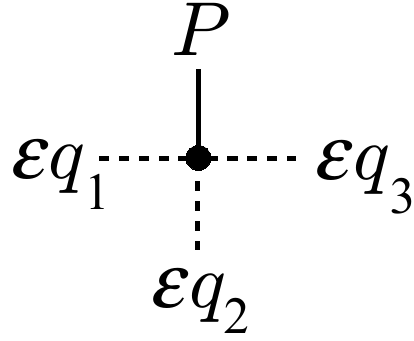}}_\sym=~0\,,\quad\qquad
    \parbox[c]{1.7cm}{\includegraphics[width=1.7cm]{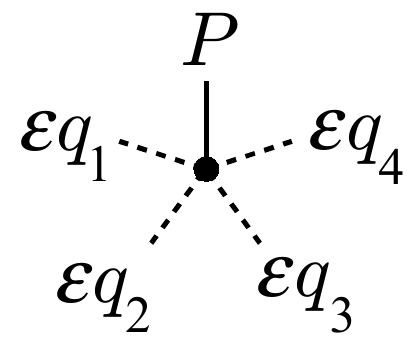}}_\sym=~0\,.
\end{equation}
This is of course simply a consequence of the $U(1)$-decoupling identity, because the sum over permutations in particular contains all cyclic permutations of the scalar. We can thus drop all diagrams that contain any of the MHV vertices in~(\ref{allptsym}). In particular, {\em this eliminates any diagrams with propagators that diverge in the collinear limit $q_i\to q$.}

Consider now vertices with two or more lines that are not vev scalars. Again, several classes of such vertices vanish:
\begin{equation}\label{moreallptsym}
    \parbox[c]{2.2cm}{\includegraphics[width=2cm]{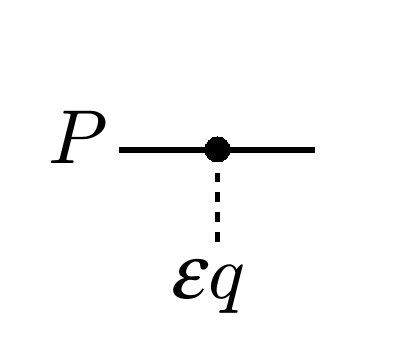}}=~0\,,\quad\qquad~~~
    \parbox[c]{.9cm}{\includegraphics[width=1cm]{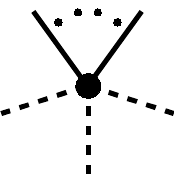}}_\sym=~0\,,\quad\qquad~~~
    \parbox[c]{1.05cm}{\includegraphics[width=.95cm]{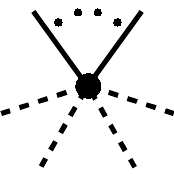}}_\sym=~0\,.
\end{equation}
Here, the first MHV vertex --- with one vev scalar and two arbitrary other lines (at least one of which is internal) --- vanishes  due to our choice of reference spinor $|q]$. Indeed, its spin factor in the numerator contains a factor of $\<P|P\!+\!\varepsilon q\>^2$ that vanishes due to the CSW prescription~(\ref{CSW}) for internal lines. The other two types of MHV vertices in~(\ref{allptsym}) vanish because they contain factors $\<q_iq_j\>$ in the numerator that go to zero in the collinear limit (while their denominators are finite after symmetrization).

The 4-point MHV vertex with two vev scalars and two other lines will be of particular importance in the following; fortunately it is very simple with our choice of reference spinor: it is only non-vanishing when the states on the two non-vev-scalar lines are conjugate to each other, and then it is simply given by
\begin{equation}\label{4pt2soft}
        \parbox[c]{2.2cm}{\includegraphics[width=2cm]{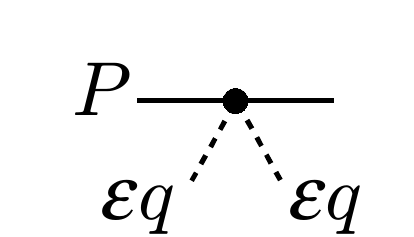}}_\sym~=~-\frac{1}{2}\<\phi_{ab}\>\<\phi^{ab}\>~=~-m^2\,.
\end{equation}
This 4-point vertex thus represents a kind of ``mass interaction vertex''. Note that~(\ref{4pt2soft}) holds even at finite $\varepsilon$; there are no higher-$\varepsilon$ corrections to this vertex due to our choice of $|q]$ as the reference spinor in the CSW expansion.

Let us now consider the final possibility of  attaching  vev scalars to an MHV vertex with at least three other lines. At most two vev scalars can attach to any one MHV vertex to get a non-vanishing diagram, so the remaining possibilities are exhausted by:
\begin{equation}\label{massvert}
    \parbox[c]{1.1cm}{\includegraphics[width=1.3cm]{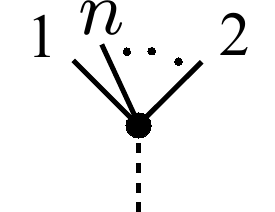}}\!\!=~\frac{m\,\<12\>}{\<1q\>\<2q\>}\times\frac{\<\ldots\>^4}{\<12\>\cdots\<n1\>}\,,\quad\qquad
    \parbox[c]{1.5cm}{\includegraphics[width=1.5cm]{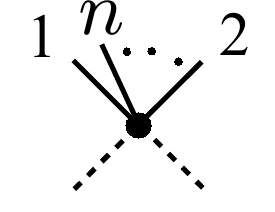}}_\sym=~\frac{m^2\<12\>^2}{\<1q\>^2\<2q\>^2}\times\frac{\<\ldots\>^4}{\<12\>\cdots\<n1\>}\,,\quad\qquad
\end{equation}
Here, the MHV numerator factors $\<\ldots\>^4$  depend on the choice of states on the solid lines; the $m$-dependence arises from the vev scalars $\<\phi_{ab}\>\phi^{ab}$ on the dashed lines.

In summary, we see that symmetrization, when acting on the MHV vertex expansion with reference vector $|q]$, not only {\em manifestly eliminates all collinear divergences}; it also considerably simplifies the amplitudes by eliminating a large number of diagrams!
\subsection{Coulomb-branch amplitudes to leading order, revisited}\label{secrev}
With the diagrammatic analysis above at hand, we can now recover some of the leading-order results of~\cite{Craig:2011ws} in a straight-forward way.
For the leading $O(m^2)$ contribution to $\< W^- \barW^+ \phi^{34} \phi^{34}\>$ in~(\ref{4ptRviol}), for example, we need to compute the  NMHV amplitude $\<g^-\phi^{\rm vev}_{\varepsilon q}\phi^{\rm vev}_{\varepsilon q}g^+\phi^{34}\phi^{34}\>_{\rm sym}$ in the massless theory. Only a single MHV  diagram can be built from the few non-vanishing MHV vertices discussed above:
\begin{equation}\label{lead4pt}
    \lim_{\varepsilon\to 0}\big\<\,g^-_1\,\phi^{\rm vev}_{\varepsilon q}\phi^{\rm vev}_{\varepsilon q}\,g^+_2\,\phi^{34}_3\, \phi^{34}_4\,\big\>_{\rm sym}~=~
     \parbox[c]{2.6cm}{\includegraphics[width=2.6cm]{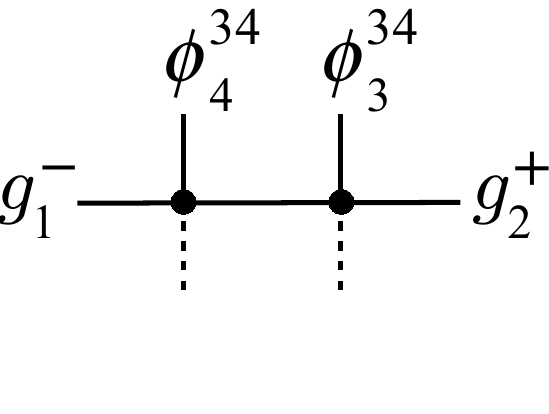}}
     ~=~ -\frac{m^2\<1^\perp|q|2^\perp]}{\<2^\perp|q|1^\perp](P_{23}^\perp)^2}\,,
\end{equation}
where  the non-vanishing contribution comes from the term $m\,\phi^{12}$ in the vev scalars~(\ref{phivev}).
The right-hand side is indeed the $O(m^2)$ contribution to~(\ref{4ptRviol}), which was computed in a more elaborate way in~\cite{Craig:2011ws}.

For the leading contribution to the $n$-point generalization of this amplitude, $\< W^-_1 \barW^+_2 \phi^{34}_3\ldots\phi_n^{34}\>$, we need to compute $\<g_1^-\phi^{\rm vev}_{\varepsilon q}\ldotss\phi^{\rm vev}_{\varepsilon q}g_2^+\phi^{34}_3\ldots \phi^{34}_n\>_{\rm sym}$ with $n\!-\!2$ vev scalars, as we can see from~(\ref{propn}). Again, only a single MHV vertex diagram contributes to this symmetrized $(2n\!-\!2)$-point N$^{n\!-\!3}$MHV amplitude, namely
\begin{equation}\label{leadnpt}
    \lim_{\varepsilon\to 0}\bigl\<\,g^-_1\,\underbrace{\phi^{\rm vev}_{\varepsilon q}\ldotss\phi^{\rm vev}_{\varepsilon q}}_{n\!-\!2
    \text{ times}}\,g^+_2\,\phi^{34}_3\ldots \phi^{34}_n\,\bigr\>_{\rm sym}
    ~=~\parbox[c]{3.1cm}{\includegraphics[width=3.1cm]{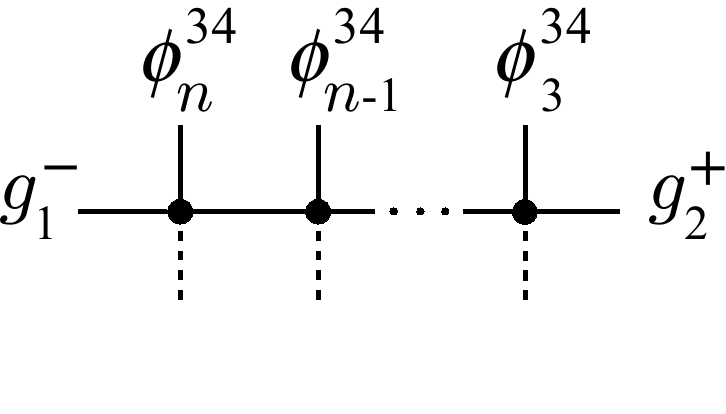}}
     ~=~ \frac{-m^{n-2}\<1^\perp|q|2^\perp]}{\<2^\perp|q|1^\perp](P_{23}^\perp)^2(P_{234}^\perp)^2\cdots(P_{23\ldots n\dash1}^\perp)^2}
    \,.
\end{equation}
To obtain this result, we express the 4-point MHV vertices with internal gluon lines as
\begin{equation}
  \parbox[c]{1.7cm}{\includegraphics[width=1.7cm]{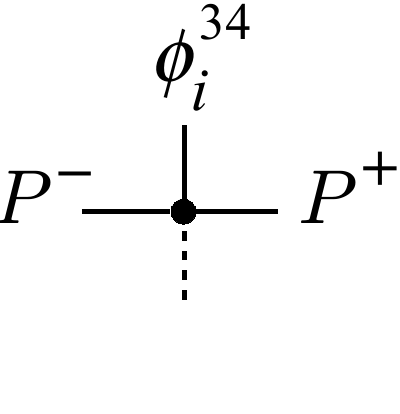}}~~=~m\frac{\<iP^-\>\<qP^-\>}{\<iP^+\>\<qP^+\>}~=~\frac{m\<q|P^-|q]}{\<q|P^-\!+\!p_i|q]}\,,
\end{equation}
where we used $P^+\!=\!-P^-\!-\!p_i$ and the CSW prescription $|P\>\!=\!P|q]$. The product over these internal MHV vertices thus collapses, yielding the result~(\ref{leadnpt}). Note that our convenient representation of the massless symmetrized soft-scalar amplitude has thus reproduced one of the results of~\cite{Craig:2011ws}  in an effortless way. We can now venture beyond the leading order.

\subsection{A full 4-point amplitude from soft limits}
To see how the full 4-point amplitude $\< W^- \barW^+ \phi^{34} \phi^{34}\>$ emerges from soft limits, we  start with  the first subleading order, $O(m^4)$. According to the proposal~(\ref{generalization}), it can be computed from a massless amplitude with four vev scalars, inserted between the gluons $g_1^-$ and $g_2^+$. Only three diagrams contribute, and they all arise from attaching a single additional 4-point vertex of
type~(\ref{4pt2soft}), \,\parbox[c]{.45cm}{\includegraphics[width=.45cm]{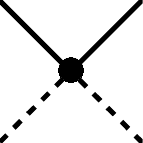}}\,, to the  diagram in~(\ref{lead4pt}):
\begin{equation}\label{subl4pt}
    \bigl\<\,g_1^-\,\phi^{\rm vev}_{\varepsilon q}\phi^{\rm vev}_{\varepsilon q}\phi^{\rm vev}_{\varepsilon q}\phi^{\rm vev}_{\varepsilon q}\,g_2^+\,\phi^{34}_3\phi^{34}_4\,\bigr\>_{\rm sym}~=~
    \parbox[c]{2.6cm}{\includegraphics[width=2.6cm]{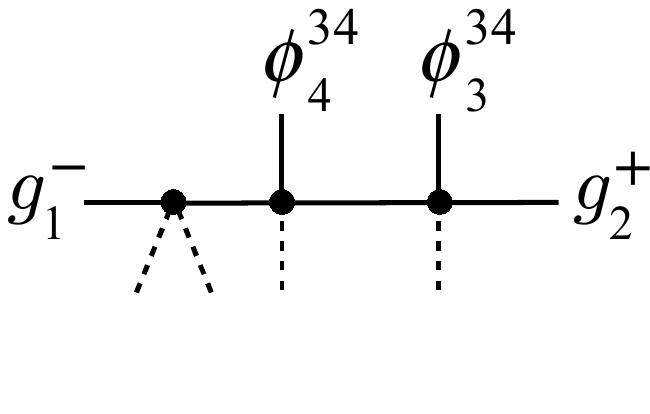}} ~+~
    \parbox[c]{2.6cm}{\includegraphics[width=2.6cm]{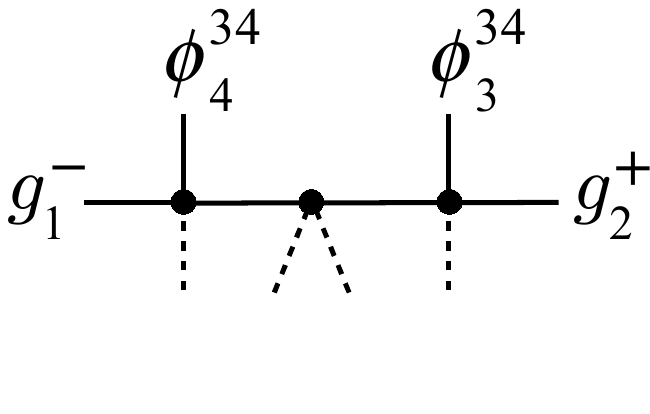}} ~+~
    \parbox[c]{2.6cm}{\includegraphics[width=2.6cm]{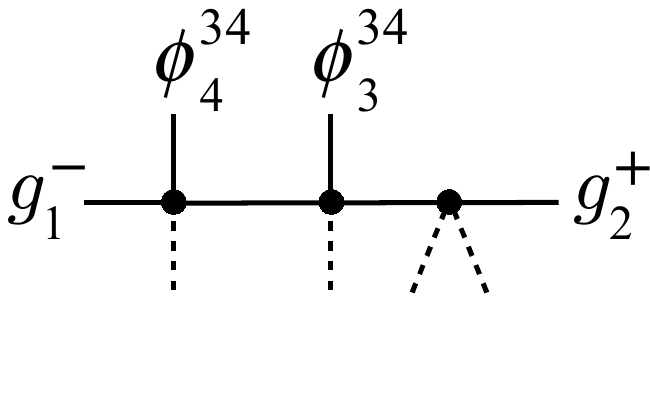}}\,.
\end{equation}
 Since 4-point vertices with two vev scalars are only non-vanishing when the particles on the remaining two lines are conjugate to each other, there are only gluons propagating on the internal lines in the diagrams of~(\ref{subl4pt}).
In the first and the last diagram, the vev scalars attach directly to an external line, so these diagrams exhibit a soft divergence in the limit $\varepsilon\to0$. Explicitly, their sum is given by
\begin{equation}
\begin{split}\label{subl4ptDIV}
    \parbox[c]{2.6cm}{\includegraphics[width=2.6cm]{subl4ptL}} ~+~
    \parbox[c]{2.6cm}{\includegraphics[width=2.6cm]{subl4ptR}}
    ~&=~\frac{m^{2}\<1^\perp|q|2^\perp]}{\<2^\perp|q|1^\perp]}
    \!\times\!\biggl[\frac{m^2}{(p_1^\perp\!+\!2\varepsilon q)^2(P_{23}^\perp)^2}
    +\frac{m^2}{(P_{23}^\perp\!+\!2\varepsilon q)^2(p_2^\perp\!+\!2\varepsilon q)^2}\biggr]\\
    ~&=~\frac{m^{2}\<1^\perp|q|2^\perp]}{\<2^\perp|q|1^\perp](P_{23}^\perp)^2}
    \!\times\!\biggl[\frac{1}{\varepsilon}\biggl(\frac{m^2}{4 \,q\!\cdot\!p_1}
    +\frac{m^2}{4 \,q\!\cdot\!p_2}\biggr)
    -\frac{m^2\, q\!\cdot\! P_{23}}{(P_{23}^\perp)^2\,q\!\cdot\! p_2}
    +O(\varepsilon)\biggr]\,.
\end{split}
\end{equation}
Here the product of MHV vertices turned into an $\varepsilon$-independent overall factor; indeed, it is obvious from~(\ref{CSW}) that  CSW spinors $|P_I\>$ are invariant under shifts of $P_I$ in the $q$ direction and thus independent of where the vev scalars attach to the diagram. Therefore diagrams that differ only by distinct insertions of 4-point MHV vertices with two soft scalars, as in~(\ref{subl4pt}), will always be accompanied by the same $\varepsilon$-independent product of the remaining MHV vertices.
We now recognize the importance of picking the special choice of $q$,~(\ref{spq}), which implies $q\!\cdot\!p_1=-q\!\cdot\!p_2$ in the two-mass case. For this choice of $q$, the coefficient of the $1/\varepsilon$ divergence in~(\ref{subl4ptDIV}) precisely cancels, and a finite unambiguous\footnote{To satisfy momentum conservation, we need to pick massless momenta $p_i^\perp$ that satisfy $\sum_i p_i^\perp\!+\!4\varepsilon q=0$. In the limit $\varepsilon\to 0$ these $p_i^\perp$ approach the projections~(\ref{decomp}) of the massive Coulomb-branch momenta. However, it is important to pick the $p_i^\perp$ in a way that they  satisfy~(\ref{spq}) also at non-vanishing $\varepsilon$ to avoid finite prescription-dependent terms in the soft-scalar amplitude.} piece remains in the $\varepsilon\to0$ limit:
\begin{equation}
    \lim_{\varepsilon\to0} \parbox[c]{2.6cm}{\includegraphics[width=2.6cm]{subl4ptL}} ~+~
    \parbox[c]{2.6cm}{\includegraphics[width=2.6cm]{subl4ptR}}
    ~=~-\frac{\<1^\perp|q|2^\perp]}{\<2^\perp|q|1^\perp](P_{23}^\perp)^2}
    \times\frac{m^{4}\,q\!\cdot\! P_{23}}{  (P_{23}^\perp)^2\,q\!\cdot\! p_2}\,.
\end{equation}
Recalling the Coulomb-branch amplitude~(\ref{4ptRviol}), we see that this finite term in the soft limit precisely recovers the first subleading contribution from the $O(m^2)$ piece of $P_{23}\!=\!P_{23}^\perp\!-\!\frac{m^2}{2q\cdot p_2}q$ in the propagator $1/(P_{23}^2\!+\!m^2)$.

To also recover the `$+m^2$' piece in the propagator, we consider the remaining second diagram in~(\ref{subl4pt}):
\begin{equation}
        \lim_{\varepsilon\to0}\parbox[c]{2.6cm}{\includegraphics[width=2.6cm]{subl4ptI}}
        ~=~\frac{\<1^\perp|q|2^\perp]}{\<2^\perp|q|1^\perp]}\times\frac{m^4}{(P_{23}^\perp)^4}\,.
\end{equation}
This diagram is finite even without invoking the special choice of $q$.
Its contribution accounts for the first subleading contribution from the `$+m^2$' piece of the Coulomb-branch propagator $1/(P_{23}^2\!+\!m^2)$ in $\< W^- \barW^+ \phi^{34} \phi^{34}\>$. Summing all three diagrams, we obtain
\begin{equation}
    \lim_{\varepsilon\to0}\bigl\<\,g_1^-\,\phi^{\rm vev}_{\varepsilon q}\phi^{\rm vev}_{\varepsilon q}\phi^{\rm vev}_{\varepsilon q}\phi^{\rm vev}_{\varepsilon q}\,g_2^+\,\phi^{34}_3\phi^{34}_4\,\bigr\>_{\rm sym}
        ~=~-\frac{\<1^\perp|q|2^\perp]}{\<2^\perp|q|1^\perp](P_{23}^\perp)^2}
    \times\frac{m^{4}\,q\!\cdot\! p_{3}}{  (P_{23}^\perp)^2\,q\!\cdot\! p_2}\,.
\end{equation}
This is indeed the $O(m^4)$ term in the expansion~(\ref{4ptexpand}) of the amplitude $\< W^- \barW^+ \phi^{34} \phi^{34}\>$.

The pattern suggested by the above computation of the first subleading contribution generalizes to all orders: the only diagrams that contribute to the massless amplitude
$\<g^-\phi^{\rm vev}_{\varepsilon q}\ldotss\phi^{\rm vev}_{\varepsilon q}g^+\phi^{34}\phi^{34}\>_{\rm sym}$ with $s$ vev scalars
arise from the leading  diagram in~(\ref{lead4pt}) by inserting additional 4-point MHV vertices \,\parbox[c]{.45cm}{\includegraphics[width=.45cm]{fourPoint}}\, with two vev-scalar lines. Inserting $v$ such additional vertices  in all possible ways on the lines $1$, $2$, and $P_{23}^\perp$ determines the symmetrized soft-scalar amplitude with $s=2v+2$ scalars as
\begin{equation}\label{sumvis}
\begin{split}
    \bigl\<\,g^-_1\,\underbrace{\phi^{\rm vev}_{\varepsilon q}\ldotss\phi^{\rm vev}_{\varepsilon q}}_{s=2v\!+\!2
    \text{ times}}\,g_2^+\,\phi^{34}_3\phi^{34}_4\,\bigr\>_{\rm sym}
    ~=\!\sum_{v_L+v_I+v_R=v}\,&
       \parbox[c]{6.6cm}{\includegraphics[width=6.6cm]{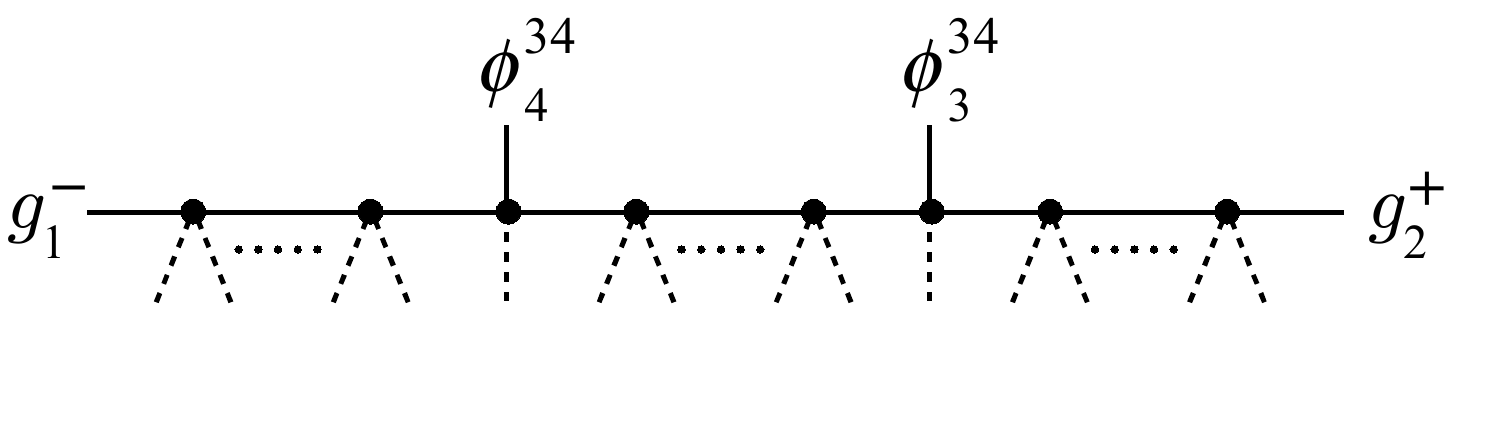}}\,.
    \\[-6ex]
    &\hskip.6cm
    \underbrace{\hskip1.15cm}_{v_L\text{ times}}\hskip.73cm
    \underbrace{\hskip1.15cm}_{v_I\text{ times}}\hskip.65cm
    \underbrace{\hskip1.15cm}_{v_R\text{ times}}
\end{split}
\end{equation}
Remarkably, although the individual terms in the sum  diverge  as $1/\varepsilon^{v_L+v_R}$ in the soft limit, their sum is finite {\em iff} we demand that $q$ satisfies~(\ref{spq}). Indeed, the sum over $v_i$'s in~(\ref{sumvis}) can be carried out analytically, and we find the very simple finite answer
\begin{equation}\label{222}
    \lim_{\varepsilon\to 0}\,\bigl\<\,g^-_1\,\underbrace{\phi^{\rm vev}_{\varepsilon q}\ldotss\phi^{\rm vev}_{\varepsilon q}}_{s=2v\!+\!2
    \text{ times}}\,g_2^+\,\phi^{34}_3\phi^{34}_4\,\bigr\>_{\rm sym}
    ~=~-\frac{m^2\<1^\perp|q|2^\perp]}{\<2^\perp|q|1^\perp](P_{23}^\perp)^2}\times\biggl(m^2\frac{q\!\cdot\! p_3}{(P_{23}^\perp)^2\,q\!\cdot\! p_2}\biggr)^{v}
    \,.
\end{equation}
The derivation of this result is presented explicitly in appendix~\ref{appA1}. Here, we just note that the crucial step that ensures the cancellation of soft divergences in~(\ref{sumvis}) is the following identity:
\begin{equation}\label{id}
    \frac{1}{\varepsilon^t}\sum_{r=0}^t \frac{(-1)^r}{r!(t-r)!}\prod_{i=1}^N\frac{1}{1+  \varepsilon r x_i}
    ~=~ \!\!\!\!\!\sum_{t_1+\ldots+t_N=t}\,\,\prod_{i=1}^N x_i^{t_i}~+~O(\varepsilon)\,.
\end{equation}
The massless amplitudes~(\ref{222})  precisely reproduce the mass expansion~(\ref{4ptexpand}) of $\< W^- \barW^+ \phi^{34} \phi^{34}\>$ to all orders! This verifies the proposal~(\ref{generalization}) for this particular amplitude. We will now turn to the $n$-point generalization of this example.

\subsection{An $n$-point amplitude from soft limits}\label{secnptSU4V}
Let us now consider the maximally $SU(4)$-violating $n$-point~(\ref{nptRviol}), $\< W^-_1 \barW^+_2 \phi^{34}_3\ldots \phi^{34}_n \>$. Drawing the lessons from our 4-point example above, it is straight-forward to generalize the analysis to this case. However, it is instructive to organize terms in a slightly different way in this all-$n$ example, as this will guide to the structure for general Coulomb-branch amplitudes.

In section~\ref{secrev}, we reproduced the leading term in this $n$-point Coulomb-branch amplitude from a single MHV vertex diagram in the massless soft-scalar amplitude,~(\ref{leadnpt}). The all-order expression~(\ref{nptRviol}) differs from this leading term simply by the replacement $(P_{2\ldots i}^\perp)^2\to P_{2\ldots i}^2\!+\!m^2$ on all propagators, which we now try to reproduce from massless soft-scalar amplitudes.

In the first subleading contribution of the 4-point example above, we saw  the  correction to the full propagator building up in two stages: The finite sum of the two soft-divergent diagrams accounted for the shift $P_{23}^\perp\to P_{23}\!=\!P_{23}^\perp\!-\!\frac{m^2}{2q\cdot p_2}q$, while the four-vertex insertion on the internal reproduced the shift `$+m^2$' to leading order. It is thus convenient to introduce the following graphical representations for the different types of `propagators':
\begin{equation}\label{props}
    \parbox[c]{1.8cm}{\includegraphics[width=1.6cm]{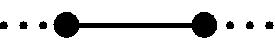}}~=~\frac{1}{(P_{I}^\perp)^2}\,, \qquad\quad
    \parbox[c]{1.8cm}{\includegraphics[width=1.6cm]{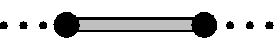}}~=~\frac{1}{P_{I}^2}\,, \qquad\quad
    \parbox[c]{1.8cm}{\includegraphics[width=1.6cm]{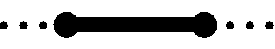}}~=~\frac{1}{P_I^2+m^2}\,.
\end{equation}

As in the 4-point case, the diagrams contributing to the massless amplitude $\<g_1^-\phi^{\rm vev}_{\varepsilon q}\!\ldotss\phi^{\rm vev}_{\varepsilon q}g_2^+\phi^{34}_3\!\!\ldots\phi^{34}_n\>_{\rm sym}$ for general $s$ arise from the leading-order $s\!=\!n\!-\!2$ diagram in~(\ref{leadnpt}) by attaching additional 4-point vertices \,\parbox[c]{.45cm}{\includegraphics[width=.45cm]{fourPoint}}\, in all possible ways to the external lines $1$ and $2$ and to all internal lines; schematically, we thus need to sum diagrams of the form
\begin{equation}
\parbox[c]{6.2cm}{\includegraphics[width=6.3cm]{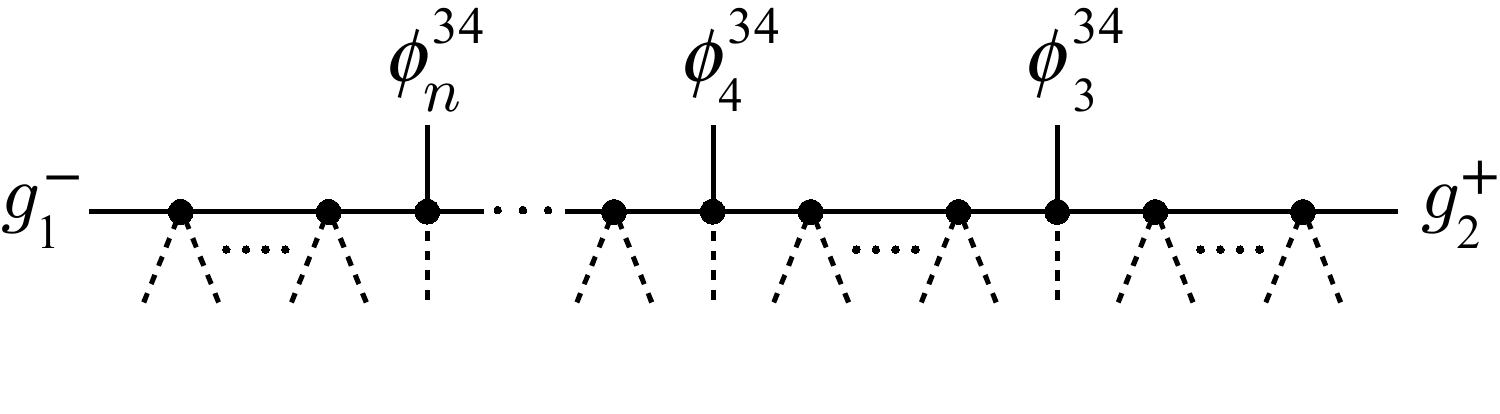}}\,.
\end{equation}

Let us first collect the contributions from soft-divergent diagrams, to all orders. Soft divergences are caused by four-point vertices
attaching directly to the external legs $1$ and $2$. We consider a fixed diagram without any such ``dangerous'' 4-point vertices, and now sum over all ways of adding $v_L$ dangerous 4-point vertices to line 1 and $v_R$ dangerous vertices to line 2.
As we show in appendix~\ref{appA2}, this sum over diagrams with soft divergence $1/\varepsilon^{v_L\!+\!v_R}$ is finite and precisely accounts for the $O(m^2)$ shift in the momenta on the external lines; it effectively replaces all internal propagators $1/(P_I^\perp)^2$ by $1/P_I^2$:
\begin{equation}
    \begin{split}\label{vlvrsum}
    \sum_{v=0}^\infty\,\sum_{v_L+v_R=v}~~\,
       &\parbox[c]{6.2cm}{\includegraphics[width=6.2cm]{propperpnpt}}
       ~~=~~\,\parbox[c]{4.1cm}{\includegraphics[width=4.1cm]{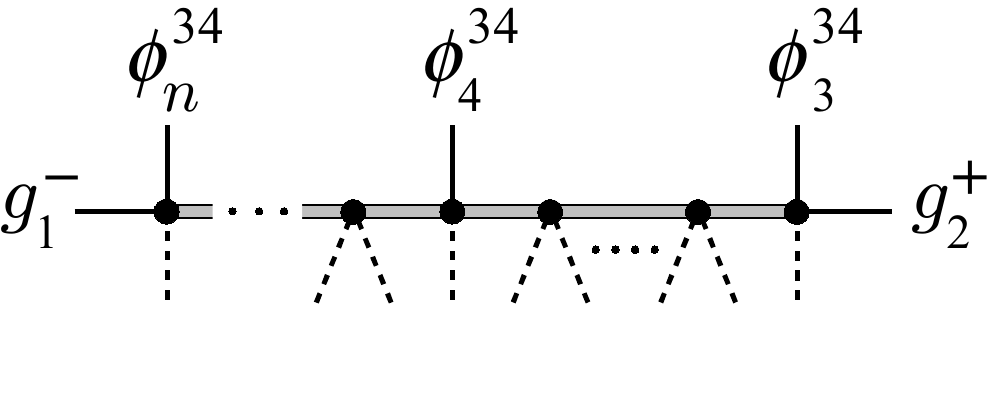}}~+~O(\varepsilon)\,.
    \\[-5.75ex]
    &\hskip.46cm
    \underbrace{\hskip1.cm}_{v_L\text{ times}}\hskip2.9cm
    \underbrace{\hskip1.cm}_{v_R\text{ times}}
\end{split}
\end{equation}
The origin of~(\ref{vlvrsum}) is again the identity~(\ref{id}).

Next, we notice that we can sum the 4-point vertices attached to each of the internal propagators:
\begin{equation}
\begin{split}\label{visum}
  \lim_{\varepsilon\to 0}\sum_{v_I=0}^\infty
  ~\,&\parbox[c]{2.5cm}{\includegraphics[width=2.5cm]{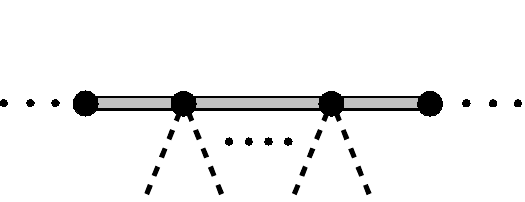}}
  ~\,=~\,\frac{1}{P_I^2}\sum_{v_I=0}^\infty\Bigl(\frac{-m^2}{P_I^2}\Bigr)^{v_I}
  ~=~\,\frac{1}{P_I^2+m^2}~=~~\parbox[c]{1.8cm}{\includegraphics[width=1.6cm]{propm2}}\,.
  \\[-3.5ex]
  &\hskip.6cm
    \underbrace{\hskip1.15cm}_{v_I\text{ times}}
\end{split}
\end{equation}
Finally, we  combine~(\ref{vlvrsum}) with~(\ref{visum}) and find
\begin{equation}
\begin{split}\label{SU4Vlaststep}
    \lim_{\varepsilon\to 0}\,\sum_{s=n-2}^\infty \,
    \bigl\<\,g_1^-\,\underbrace{\phi^{\rm vev}_{\varepsilon q}\ldotss\phi^{\rm vev}_{\varepsilon q}}_{s \text{ times}}\,g_2^+\,\phi^{34}_3\ldots\phi^{34}_n\,\bigr\>_{\rm sym}~&=~\parbox[c]{3cm}{\includegraphics[width=3cm]{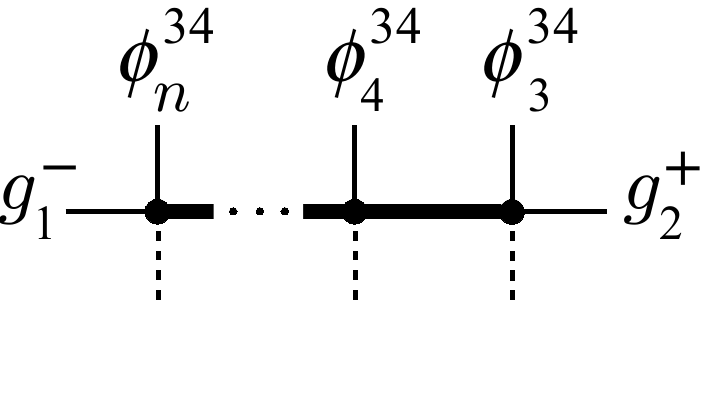}}\\[-2ex]
    &=~\frac{-m^{n-2}\<1^\perp|q|2^\perp]}{\<2^\perp|q|1^\perp](P_{23}^2+m^2)\cdots(P_{23\ldots n\dash 1}^2+m^2)}\,.
\end{split}
\end{equation}
This proves~(\ref{propn}) to all orders.

\setcounter{equation}{0}
\section{CSW-like expansion for two-mass amplitudes}\label{CSW2}
Our analysis in section~\ref{secnptSU4V} suggests a simple CSW-like expansion for Coulomb-branch amplitudes with 2 adjacent massive lines. Indeed, as shown above, one can resum the contributions from the 4-point vertices of type \,\parbox[c]{.45cm}{\includegraphics[width=.45cm]{fourPoint}}\, in the proposal~(\ref{generalization}) and obtain an expansion of the Coulomb-branch amplitude in terms of diagrams that contain the remaining vertices, connected by massive propagators on the massive lines.

Let us state the diagrammatic rules for this CSW-like expansion explicitly: The two massive lines (say lines $1$ and $2$) are connected by a string of massive scalars propagators with momenta $P_{2..i}$,
while all other internal lines carry massless scalar propagators of momenta $P_{i..j}$:
\begin{equation}
\begin{split}
    &\hskip.45cm P_{2..i}\hskip 5.13cm P_{i..j}\\[-2.9ex]
     &\parbox[c]{1.8cm}{\includegraphics[width=1.6cm]{propm2}}~=~\frac{1}{P_{2..i}^2+m^2}\,, \qquad\qquad
     \parbox[c]{1.8cm}{\includegraphics[width=1.6cm]{propperp}}~=~\frac{1}{P_{i..j}^2}\, \qquad\qquad\text{with }2<i<j\,.
\end{split}
\end{equation}
 The vertices that only connect massless particles are conventional $n$-point MHV vertices. In the language of superamplitudes, they are given by
 \begin{equation}
    \parbox[c]{2.2cm}{\includegraphics[width=1.5cm]{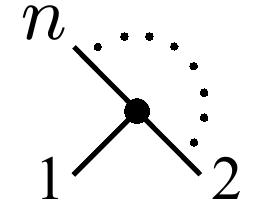}}~=~\frac{\delta^{(8)}\big(|i\>\eta_{ia}\big)}{\<12\>\cdots\<n1\>}\,.
 \end{equation}
 As usual, the CSW prescription~(\ref{CSW}) for internal lines is understood.

 There are three types of  vertices that contain massive lines (which we take to be lines $1$ and $2$):
\begin{equation}
\begin{split}\label{massvertcsw}
\parbox[c]{1.7cm}{\includegraphics[width=1.5cm]{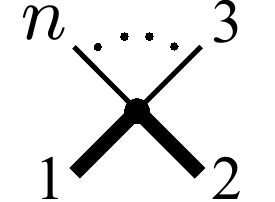}}~&=~\frac{\delta^{(8)}\big(|i^\perp\>\eta_{ia}\big)}{\<1^\perp2^\perp\>\cdots\<n\,1^\perp\>}\,,\\
    \parbox[c]{1.7cm}{\includegraphics[width=1.5cm]{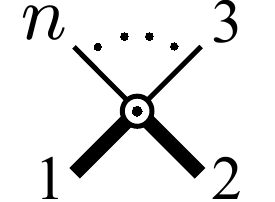}}~&=~\frac{m\<1^\perp2^\perp\>}{\<1^\perp q\>\<2^\perp q\>}\times\frac{\delta^{(4)}_{12}\big(|i^\perp\>\eta_{ia}\big)\,\delta^{(2)}_{34}\big(\<qi^\perp\>\eta_{ia}\big)
    +\delta^{(4)}_{34}\big(|i^\perp\>\eta_{ia}\big)\,\delta^{(2)}_{12}\big(\<qi^\perp\>\eta_{ia}\big)
    }{\<1^\perp2^\perp\>\cdots\<n\,1^\perp\>}\,,\\
\parbox[c]{1.7cm}{\includegraphics[width=1.5cm]{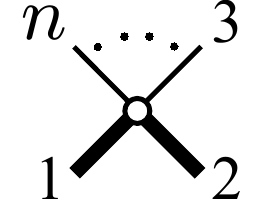}}~&=~\frac{m^2\<1^\perp2^\perp\>^2}{\<q1^\perp\>^2\<q2^\perp\>^2}\times\frac{\delta^{(4)}\big(\<qi^\perp \>\eta_{ia}\big)}{\<1^\perp2^\perp\>\cdots\<n\,1^\perp\>}\,.
\end{split}
\end{equation}
The last two vertices are derived from MHV vertices with one and two vev-scalar insertions, respectively, as is evident from comparing~(\ref{massvertcsw}) to~(\ref{massvert}). The subscripts on the Grassmann $\delta$-functions  indicate  which of the two $SU(2)$ factors of $Sp(4)$ the $\delta$-function `lives in'.
We did not separate out massive and massless lines in the arguments of the $\delta$-functions, and instead used $|i^\perp\>\equiv|i\>$ for massless external momenta.
Massless states are projected out of the  $\delta$-function using the Grassmann differential operators presented in~\cite{Bianchi:2008pu}. Massive states are projected out using the differential operator of the corresponding massless particle as specified by the dictionary in Table~\ref{tabstates}.
All of the above $n$-point vertices exist for any $n\geq 3$.

It is worth pointing out a key difference between this CSW-like expansion and the MHV vertex expansion for massless amplitudes. In the massless case, all vertices are actual amplitudes; indeed, any MHV amplitude only receives contributions from the corresponding MHV vertex. In the massive CSW-like expansion, however, the vertices are not actual massive amplitudes. In fact, Coulomb-branch amplitudes whose external states correspond to one of the vertices in~(\ref{massvertcsw}) can receive additional contributions from diagrams with more than one vertex. This is illustrated in the examples below.

\subsubsection*{Relaxation of the $q$-constraint}
Up to now we have always insisted that $q$  must satisfy the condition~(\ref{spq}). This was necessary for the soft divergences to cancel in the massless vev-scalar amplitudes. In the CSW-like expansion above, however, there are no soft-divergent diagrams left. These have already been summed into massive propagators. Also, if the CSW-like expansion is valid for any amplitude for which $q$  satisfies the constraint~(\ref{spq}), then it is also valid for the subamplitudes of all factorization channels of this amplitude. However, it is easy to see that the `special $q$' of the parent amplitude does not imply the same constraint on its subamplitudes. It follows that the {\em CSW-like expansion with the diagrammatic rules given above is valid for any $q$}, not just for $q$'s that satisfy~(\ref{spq}).
 We now test this CSW-like expansion in various examples.
\subsubsection*{Examples}
As a warm-up, consider the maximally $SU(4)$-violating amplitude $\< W^-_1 \barW^+_2 \phi^{34}_3\ldots \phi^{34}_n \>$ that we computed from soft limits in section~\ref{secnptSU4V}. Only one diagram contributes, and we get
\begin{equation}
\begin{split}\label{again}
\bigl\< W^-_1 \barW^+_2\, \phi^{34}_3\ldots \phi^{34}_n \bigr\>
~&=~~\parbox[c]{3.5cm}{\includegraphics[width=3.5cm]{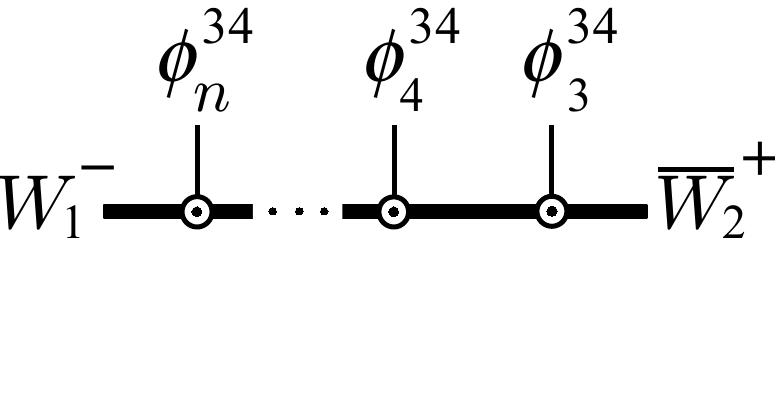}}
    ~~=~\frac{-m^{n-2}\<1^\perp|q|2^\perp]}{\<2^\perp|q|1^\perp](P_{23}^2\!+\!m^2)\cdots(P_{23\ldots n\dash1}^2\!+\!m^2)}\,.
\end{split}
\end{equation}
This is of course simply the computation~(\ref{SU4Vlaststep}), in a slightly different diagrammatic language. Note, however, that~(\ref{again}) is valid for any $q$, not just for $q$'s satisfying~(\ref{spq}).

Let us now consider a slightly less trivial example: the UHV (ultra-helicity-violating) amplitudes of the form $\< W^- \barW^+ g^+\ldots g^+\>$. Let us start at 3-point, $\< W^- \barW^+g^+\>$. In the massless limit, this 3-point amplitude is anti-MHV and thus not directly captured by the MHV vertex expansion. In the Lagrangian approach to the MHV vertex expansion one can argue for its presence from dimensional regularization~\cite{Ettle:2007qc}. In our massive CSW-like expansion, however, this amplitude is directly computable from the last vertex in~(\ref{massvertcsw})! We find
\begin{equation}\label{WWg}
\begin{split}
\bigl\< W_1^- \barW_2^+\, g^+_3\bigr\>~=~
\parbox[c]{2.3cm}{\includegraphics[width=2.3cm]{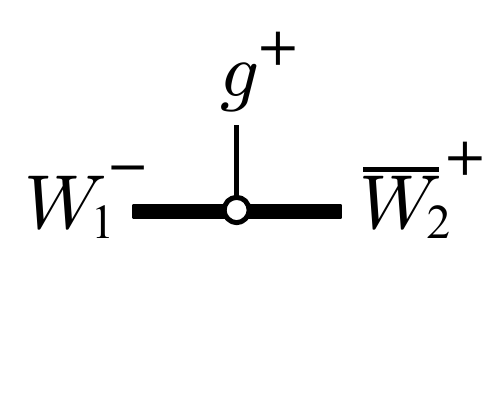}}~=~
\frac{m^2\<q1^\perp \>^2\<1^\perp 2^\perp\>}{\< q2^\perp\>^2\<2^\perp3\>\<3\,1^\perp\>}~=~\frac{[2^\perp3]^3}{[1^\perp2^\perp][3\,1^\perp]}\,.
\end{split}
\end{equation}
Although the vertex in this diagram naively goes as $\sim m^2$, the amplitude is actually $O(1)$ due to `special massive kinematics': the angle brackets $\<1^\perp 2^\perp\>$, $\<2^\perp 3\>$, and $\<3 1^\perp \>$ in this amplitude go as $\sim\! m^2$ so that they vanish in the massless limit.\footnote{It is important that we relax the constraint~(\ref{spq}) on $q$ in this computation to avoid $0/0$ ambiguities.} The rewriting on the right-hand side makes the anti-MHV nature of this amplitude manifest. Note that we did not impose the constraint~(\ref{spq}) at any stage in~(\ref{WWg}).

At 4-point, two diagrams contribute to the UHV amplitude:
\begin{equation}
\bigl\< W_1^- \barW_2^+\, g^+_3\,g_4^+\bigr\>
~=~
\parbox[c]{2.3cm}{\includegraphics[width=2.3cm]{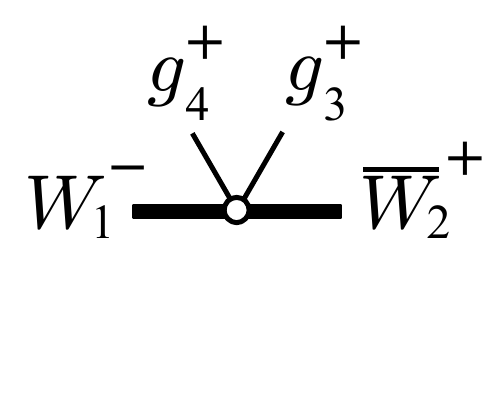}}
~+~\parbox[c]{2.75cm}{\includegraphics[width=2.75cm]{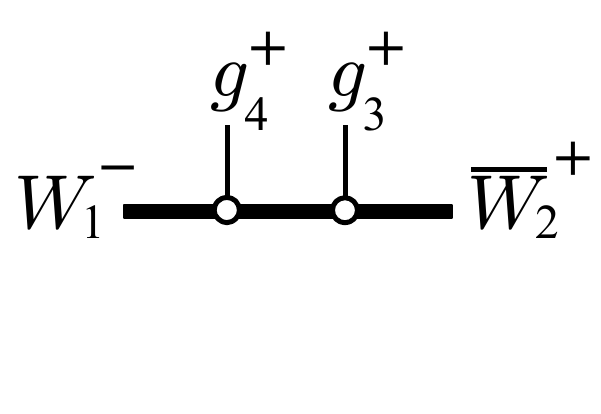}}
 ~=~
-\frac{m^2\<q1^\myperp\>^2[34]}{\<q2^\myperp\>^2\<34\>(P_{23}^2+m^2)}
\,.
\end{equation}
It takes some spinor gymnastics to combine the two diagrams into the (known) expression on the right-hand side; however, the assumption of special $q$~(\ref{spq}) is again not needed, just as we argued on general grounds above.

The general $n$-point amplitude $\< W^- \barW^+ g^+\ldots g^+\>$ can be expanded as
\begin{equation}\label{WWn}
\bigl\< W_1^- \barW_2^+\, g_3^+\ldots g_n^+\bigl\>~=~~
   \parbox[c]{2.5cm}{\includegraphics[width=2.5cm]{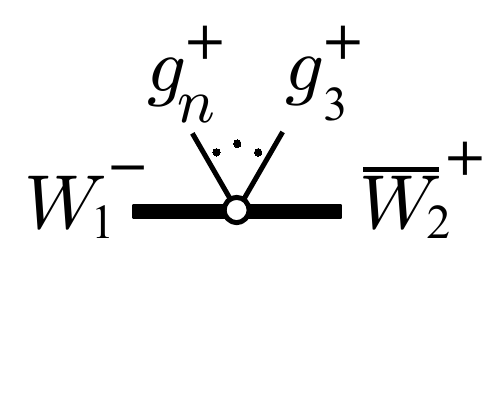}}
~~+~~\sum_{i=3}^{n-1}~\,\parbox[c]{3.75cm}{\includegraphics[width=3.75cm]{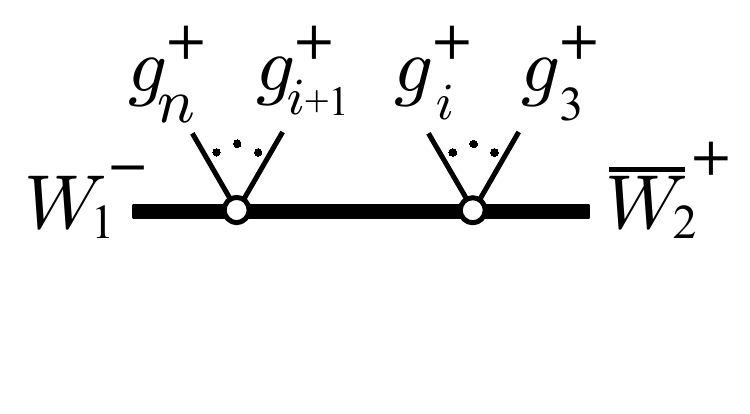}}
    ~~+~\ldots\,,
\end{equation}
with at least one positive-helicity gluon on each vertex. The `$+\ldots$' above contain further diagrams with more vertices, the final one being a string of $n\!-\!2$ vertices with one positive-helicity gluon each. To evaluate this for any $n$,
we first compute the leading vertex. It is given by
\begin{equation}
    \parbox[c]{2.5cm}{\includegraphics[width=2.5cm]{cswnptmpsdiag1}}
    ~=~\frac{m^2\<q1^\perp\>^2\<1^\perp 2^\perp\>}{\<q2^\perp\>^2\<2^\perp 3\>\<34\>\cdots \<n1^\perp\>}\,.
\end{equation}
Then we notice that successive diagrams in the sum~(\ref{WWn}) arise by `splitting' a vertex as
\begin{equation}\label{split}
       \parbox[c]{2.5cm}{\includegraphics[width=2.5cm]{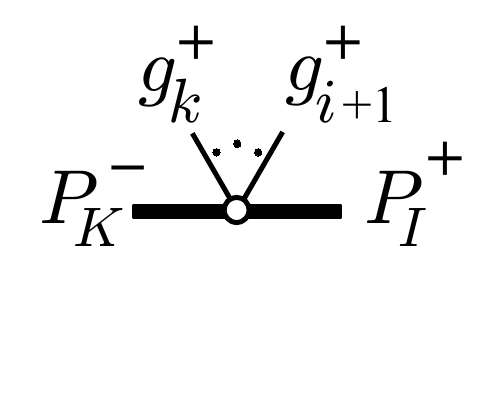}}
       ~~\longrightarrow~~
       \parbox[c]{3.75cm}{\includegraphics[width=3.75cm]{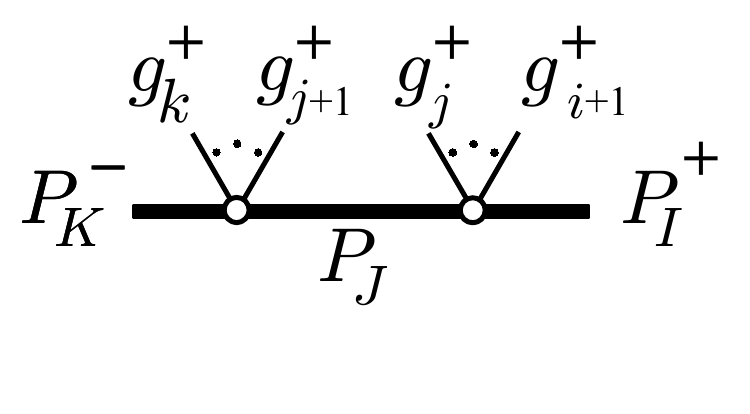}}\,,
\end{equation}
where we denoted $P_J\equiv P_{2..j}$.
Each such split introduces a relative factor of
\begin{equation}
    -\frac{m^2\,\,\<j,j\!+\!1\>\<P_IP_J\>\<P_JP_K\>}{(P_J^2+m^2)\<j\,P_J\>\<P_J,j\!+\!1\>\<P_IP_K\>}\,.
\end{equation}
It follows that we can write the full amplitude as
\begin{equation}\label{WWalln}
    \bigl\< W_1^- \barW_2^+\, g_3^+\ldots g_n^+\bigl\>~=~-
    \frac{m^2\<q1^\perp\>^2}{\<q2^\perp\>^2\<2^\perp 3\>\<34\>\cdots \<n1^\perp\>}\times \big\<2^\perp\big|\prod_{j=3}^{n-1}\biggl[1-\frac{m^2|P_J\>\<j,j\!+\!1\>\<P_J|}{(P_J^2+m^2)\<P_J,j\>\<j\!+\!1,P_J\>}\biggr]
    \big|1^\perp\big\>\,.
\end{equation}
This should be compared to the  known expression for this amplitude~\cite{Ferrario:2006np,Craig:2011ws,Boels:2011zz} (see also~\cite{Forde:2005ue,Rodrigo:2005eu}),
\begin{equation}\label{WWallnceks}
    \bigl\< W_1^- \barW_2^+\, g_3^+\ldots g_n^+\bigl\>~=~
    -\frac{m^2\<q\,1^\myperp\>^2\,}{\<q\,2^\myperp\>^2\,\<34\>\<45\>\cdots \<n\!-\!1,n\>(P_{n1}^2+m^2)}\times
    \bigl[3\big|\prod_{j=3}^{n-2}\biggl[1+\frac{P_{J}|j\!+\!1\>[j\!+\!1|}{P_{J}^2+m^2}\biggr]\big|n\big]
    \,.
\end{equation}
We have  confirmed  the equivalence of these two expressions  numerically for $n=3,4,5,6$. The form~(\ref{WWallnceks}) is arguably slightly simpler than~(\ref{WWalln}); still, it is remarkable that a relatively simple expression for this amplitude came directly out of the CSW-like diagrammatic rules, with no need to solve a BCFW recursion relation to all $n$.

\setcounter{equation}{0}
\section{General Coulomb-branch amplitudes}\label{secgeneral}
\subsection{Setup}
We now extend our analysis to general Coulomb-branch amplitudes, with arbitrary mass distributions on the external lines.
We consider a breaking $U(N)\to\prod_k U(M_k)$ of the gauge group. Though not essential for our analysis, we will stick to a breaking $SU(4)\to Sp(4)$ of the R-symmetry group. The  scalars  then have vevs
\begin{equation}
    \<(\phi^{12})_{A}{}^{B}\> = \<(\phi^{34})_A{}^{B}\big\>  = v_k\delta_{A}{}^{B}\,.
\end{equation}
The proportionality constant $v_k$ depends on the gauge group $U(M_k)$ of the index $A$; it can be interpreted as the position of the stack of $M_k$ D-branes in 6 dimensions.
Bifundamental particles $X_A{}^B$ whose color indices $A$ and $B$ belong  to distinct subgroups $U(M_k)$ and $U(M_{k'})$, respectively,  acquire a mass
\begin{equation}
    m_X~=~v_{k'}-v_k\,.
\end{equation}
Again, this is natural from a higher-dimensional perspective: it is the mass of a string stretched between two stacks of D-branes at positions $v_k$ and $v_{k'}$.

\begin{figure}
\begin{center}
\includegraphics[width=3.5cm]{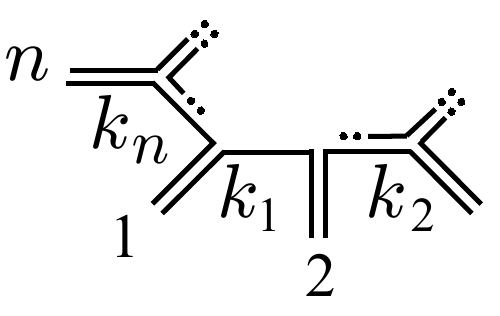}
\hskip2cm
\includegraphics[width=3.5cm]{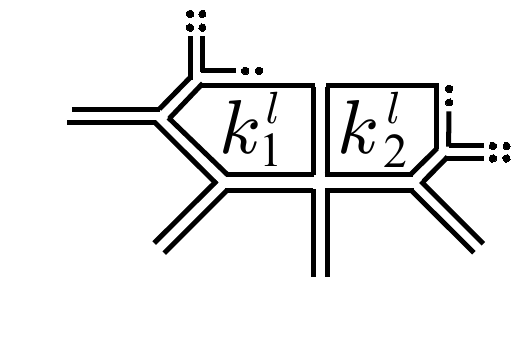}
\end{center}
\vskip-.3cm
\centerline{~~~~(a)\hskip5.3cm(b)}
\caption{On the Coulomb-branch, the gauge group is spontaneously broken as $U(N)\to \prod_k U(M_{k})$. In a double-line notation for diagrams of Coulomb-branch amplitudes, we associate a gauge group $U(M_{k})$ with each connected line.\quad(a):~The gauge group of the line connecting particle $i$ to $i\!+\!1$ is identified by $k_i$.\quad (b):~In planar $L$-loop diagrams, which we discuss in section~\ref{secloop}, the groups corresponding to internal loop lines are identified by $k^l_1,k^l_2,\ldots,k^l_L$.
}
\label{figDL}
\end{figure}

It is instructive to think of the amplitude in a double-line notation, and associate a group $U(M_k)$ with each line. Let us denote the group associated with the line connecting particle $i$ to $i+1$ by $U(M_{k_i})$. See Figure~\ref{figDL}(a).
Particle $i$ then has mass $m_i\!=\!v_{k_i}\!-\!v_{k_{i-1}}$.
For the case of arbitrary $n$-point Coulomb-branch amplitudes $\<X_1X_2\ldots X_n\>$ with arbitrary masses (subject to the Coulomb-branch constraint $\sum m_i=0$), the proposal~(\ref{generalization}) then generalizes to
\begin{equation}\label{generalization2}
    \bigl\<\,X_1X_2\,\ldots\,X_n\bigr\>~~=~~\lim_{\varepsilon\to 0}\,\sum_{s=0}^\infty~\sum_{s_1+\ldots+s_n=s} \,\bigl\<\,Y_1\,
    \underbrace{\phi^{\rm vev}_{\varepsilon q}\ldotss\phi^{\rm vev}_{\varepsilon q}}_{s_1\text{ times}}\,Y_2\,
    \underbrace{\phi^{\rm vev}_{\varepsilon q}\ldotss\phi^{\rm vev}_{\varepsilon q}}_{s_2\text{ times}}\,Y_3\,
    \ldots\,Y_n
        \underbrace{\phi^{\rm vev}_{\varepsilon q}\ldotss\phi^{\rm vev}_{\varepsilon q}}_{s_n\text{ times}}\,
    \,\bigr\>_{\rm sym}\,.
\end{equation}
Here, the particles $X_i$ in the Coulomb-branch amplitude have mass $m_{i}=v_{k_i}\!-\!v_{k_{i-1}}$, and the $Y_i$ are the massless particles  corresponding to $X_i$ as detailed in Table~\ref{tabstates}. For example, a transverse vector boson in the Coulomb-branch amplitude $X_i=W_i^\pm$ corresponds to a gluon $Y_i=g_i^\pm$ in the massless amplitude. The $s_i$ vev scalars  between  lines $Y_i$ and $Y_{i+1}$ are given by $\phi^{\rm vev}=v_{k_i}(\phi^{12}\!+\!\phi^{34})$. The symmetrization in~(\ref{generalization2}) acts separately on each of the insertions of  $s_i$ vev scalars. We now argue that the proposal~(\ref{generalization2}) is well-defined, \ie the sum on the right-hand side is free of collinear and soft divergences under the assumption that $q$ satisfies~(\ref{spq}).

\subsection{Collinear and soft divergences}

The analysis of collinear divergences in section~\ref{seccoll2} still applies to this more general case; indeed, collinear divergences only arise from adjacent scalars, so the vanishing of the diagrams~(\ref{allptsym}) is enough to ensure collinear finiteness for each symmetrized amplitude on the right-hand side of~(\ref{generalization2}).

The generalization of the cancellation of soft divergences is less trivial. In fact, the individual symmetrized amplitudes on the right-hand side of~(\ref{generalization2}) are {\em not} finite in the limit $\varepsilon\to 0$; soft divergences only cancel in the sum over all $s_i$ with fixed $\sum{s_i}=s$\,. Let us illustrate how this works for $s=2$. In the MHV vertex expansion with reference spinor $|q]$, all soft divergences again arise from two vev scalars attaching directly to an external line $i$ via a $4$-point vertex. For each line $i$, three different 4-point vertices occur in the sum~(\ref{generalization2}), depending on how the two vev scalars are distributed:
\begin{equation}
\begin{split}
          \parbox[c]{1.2cm}{\includegraphics[width=1.3cm]{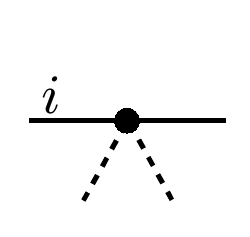}}_\sym~=~-v_{k_i}^2\,, \qquad
          \parbox[c]{1.2cm}{\includegraphics[width=1.3cm]{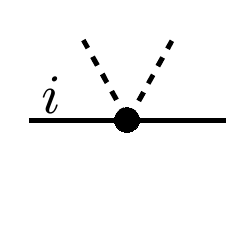}}_\sym~=~-v_{k_{i-1}}^2\,,  \qquad
          \parbox[c]{1.3cm}{\includegraphics[width=1.3cm]{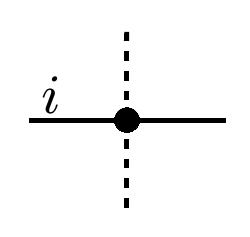}}~=~2v_{k_i}v_{k_{i-1}}\,.
\end{split}
\end{equation}
As in~(\ref{4pt2soft}), these values of the 4-point vertices are exact with no $O(\varepsilon)$ corrections.
It is convenient to introduce a diagrammatic notation for the sum over these three vertices:
\begin{equation}\label{4pt2softAll}
\parbox[c]{1.3cm}{\includegraphics[width=1.3cm]{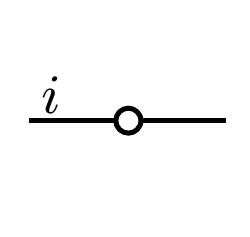}}~~\equiv~~
\parbox[c]{1.2cm}{\includegraphics[width=1.3cm]{4pt2softR}}_\sym+~
\parbox[c]{1.2cm}{\includegraphics[width=1.3cm]{4pt2softL}}_\sym+~
\parbox[c]{1.3cm}{\includegraphics[width=1.3cm]{4pt2softlr}}~=~
-(v_{k_i}\!-v_{k_{i-1}})^2~=~-m_i^2\,.
\end{equation}
We can now compute the divergent piece in the sum~(\ref{generalization2}) for $s=2$. It is given by
\begin{equation}\label{seq2}
    \sum_{i=1}^n
    ~~\parbox[c]{2.5cm}{\includegraphics[width=2.5cm]{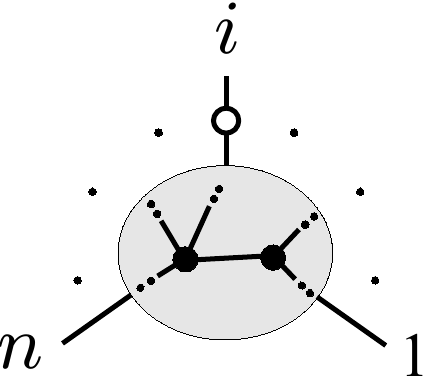}}~~
    ~~=~~\frac{1}{\varepsilon}\biggl[\,\sum_{i=1}^n\frac{-m_i^2}{4\,q\!\cdot\! p_i}~\biggr]\times\big\<Y_1Y_2\ldots Y_n\big\>~+~O(\varepsilon^0)\,.
\end{equation}
Here, the `blob' represents a sum over all MHV vertex diagrams contributing to the massless amplitude $\<Y_1Y_2\ldots Y_n\>$. The $O(\varepsilon^0)$ piece arises because the amplitude blobs actually differ by an $O(\varepsilon)$ piece depending on which external line carries the two vev scalars. As in the case of 2 massive lines, these $O(\varepsilon^0)$ pieces are crucial to get the correct massive propagator; but for now we simply observe that the soft-divergent term in~(\ref{seq2}) cancels if and only if we impose the special-$q$ constraint~(\ref{spq}). Note that the soft divergences do not cancel individually for each symmetrized amplitude on the right-hand side of~(\ref{generalization2}); indeed, each vertex~(\ref{4pt2softAll}) mixes different amplitudes on the right-hand side of~(\ref{generalization2}) as the vev scalars are distributed differently with respect to line $i$. The cancellation thus only occurs after summing all amplitudes in the $s=2$ contribution of~(\ref{generalization2}).

Proving finiteness in the soft limit to all orders for any Coulomb-branch amplitude requires careful bookkeeping and
a tedious combinatorial analysis. In the end, it can be reduced an iterated application of the identity~(\ref{id}). We present the detailed derivation in appendix~\ref{appA3}. Here we simply state the result: Summing over all collinear divergent diagrams in all amplitudes on the right-hand side of~(\ref{generalization2}) has the effect of turning all internal propagators  $1/(P^\perp_I)^2$ in the diagrams of these amplitudes into $1/P_I^2$. Schematically, using the graphical notation for propagators introduced in~(\ref{props}), we have
\begin{equation}\label{perptoprop}
\sum_{v=0}^\infty~\sum_{v_1+..+v_n=v}\hskip-1cm\parbox[c]{5cm}{\includegraphics[width=5cm]{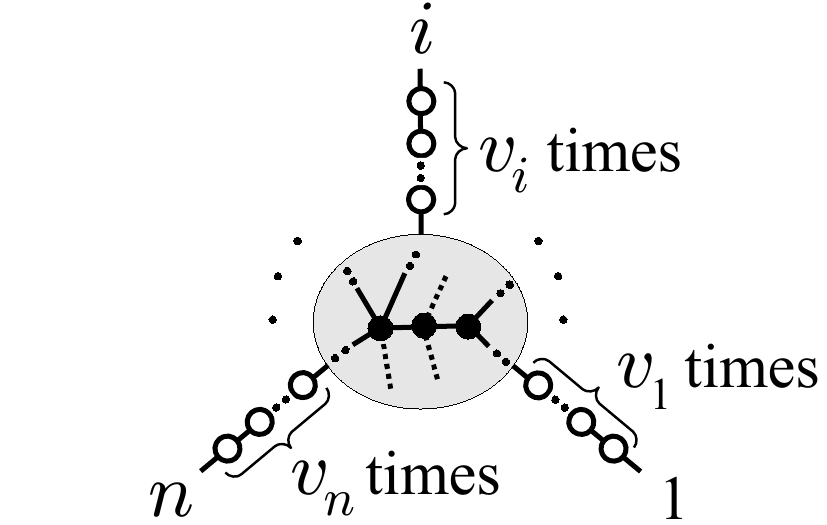}}~~~~~=~~~~~
\parbox[c]{3cm}{\includegraphics[width=3cm]{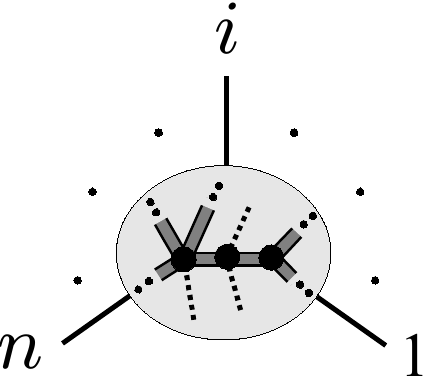}}\,.
\end{equation}
Here, the inner `blobs' contain all diagrams in the MHV vertex expansions of the amplitudes on the right-hand side of~(\ref{generalization2}) that are free of soft divergences; the soft divergences are taken into account by the $v_i$ `dangerous' 4-point vertices that we are attaching to each external line $i$. The sum in~(\ref{perptoprop}) is finite for fixed $v$; summing over $v$ then builds up the shifts $P^\perp_I\to P_I$ in the propagators as a geometric sum. The identity~(\ref{perptoprop}) is the natural generalization of~(\ref{vlvrsum}).

The identity~(\ref{perptoprop}) already establishes the soft finiteness of the proposal~(\ref{generalization2}). Still, let us press on and also sum the (finite) contributions from 4-point vertices with two vev scalars attaching to internal propagators. Just as in the case of two massive lines,~(\ref{visum}), the contribution of all such vertices amounts to a shift of the propagator that gives it the correct mass dependence for a Coulomb-branch amplitude:
\begin{equation}
\begin{split}
    \sum_{v_I=0}^\infty~~&
    \parbox[c]{3cm}{\includegraphics[width=3cm]{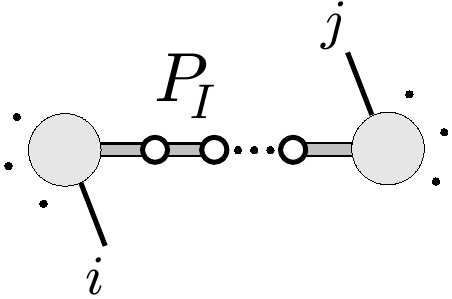}}
    ~\,=~\,\frac{1}{P_I^2}\sum_{v_I=0}^\infty\Bigl(\frac{-m_I^2}{P_I^2}\Bigr)^{v_I}
  ~=~\,\frac{1}{P_I^2+m_I^2}
    \qquad\text{ with }~~ m_I\equiv v_{k_i}\!-\!v_{k_j}=\sum_{i\in I}m_i\,,  \\[-8.5ex]
  &\hskip.85cm
    \underbrace{\hskip1.15cm}_{v_I\text{ times}}\\[-2ex]
    \phantom{A}
\end{split}
\end{equation}
where we used the same notation for the sum over the three types of 4-point vertices with two vev scalars as in~(\ref{4pt2softAll}).

\setcounter{equation}{0}
\section{CSW-like expansion for general  Coulomb-branch amplitudes}\label{CSWgen}
We now present a CSW-like expansion for general Coulomb-branch amplitudes. From the analysis in section~\ref{secgeneral}, we can immediately deduce the diagrammatic rules of this expansion.

\nopagebreak

The propagators are conventional massive scalar propagators:
\begin{equation}\label{propcsw}
        \parbox[c]{1.8cm}{\includegraphics[width=1.6cm]{propm2}}~=~\frac{1}{P_I^2+m_I^2}\,\qquad \text{with  }~m_I=\sum_{i\in I}m_i\,.
\end{equation}
This of course includes massless scalar propagators as a special case when $\sum_{i\in I}m_i\!=\!0$.

There are three types of vertices in the expansion. The first vertex is the conventional MHV vertex, with perp'ed spinors:
\begin{equation}\label{vert1}
    \parbox[c]{2.2cm}{\includegraphics[width=1.5cm]{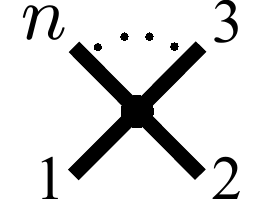}}~=~\frac{\delta^{(8)}\big(|i^\perp\>\eta_{ia}\big)}{\<1^\perp2^\perp\>\cdots\<n^\perp1^\perp\>}\,.
\end{equation}
The vertex that originates in the construction above from an MHV vertex with one insertion of the vev scalar is given by
\begin{equation}\label{vert2gen}
        \parbox[c]{1.7cm}{\includegraphics[width=1.5cm]{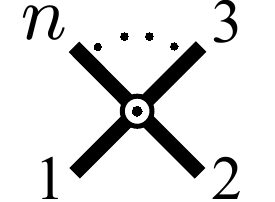}}~=~\biggl[\sum_{i}\frac{m_i\<1^\perp i^\perp\>}{\<1^\perp q\>\<i^\perp q\>}\biggr]\times\frac{\delta^{(4)}_{12}\big(|i^\perp\>\eta_{ia}\big)\,\delta^{(2)}_{34}\big(\<qi^\perp\>\eta_{ia}\big)
    +\delta^{(4)}_{34}\big(|i^\perp\>\eta_{ia}\big)\,\delta^{(2)}_{12}\big(\<qi^\perp\>\eta_{ia}\big)
    }{\<1^\perp2^\perp\>\cdots\<n^\perp1^\perp\>}\,.
\end{equation}
Note that we expressed the kinematic prefactor in terms of the masses $m_i$, not in terms of the expectation values $v_k$ of the scalar fields. This form makes it manifest that the vertex is invariant under a collective shift of all vevs, $v_k\to v_k+\delta v$.
The kinematic prefactor in~(\ref{vert2gen}) seems to single out line $1$ as special; however, the vertex does not depend on which line we single out. Indeed, using $\sum_im_i=0$, we can replace $|1^\perp\>$ by any other spinor $|X\>$:
\begin{equation}\label{vert2}
    \sum_{i}\frac{m_i\<1^\perp i^\perp\>}{\<1^\perp q\>\<i^\perp q\>}~=~
    \sum_{i}\frac{m_i\<1^\perp i^\perp\>}{\<1^\perp q\>\<i^\perp q\>}+\Bigl(\sum_i m_i\Bigr)\frac{\<X 1^\perp\>}{\<1^\perp q\>\<X q\>}
    ~=~\sum_{i}\frac{m_i\<X i^\perp\>}{\<X q\>\<i^\perp q\>}\,.
\end{equation}
Finally, there is a third vertex in the CSW-like expansion, which originates from two vev scalars attaching in all possible ways:
\begin{equation}\label{vert3}
        \parbox[c]{1.7cm}{\includegraphics[width=1.5cm]{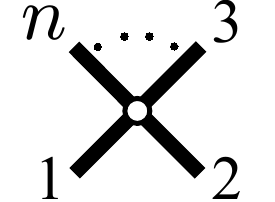}}~=~\biggl[\sum_{i}\frac{m_i\<1^\perp i^\perp\>}{\<1^\perp q\>\<i^\perp q\>}\biggr]^2\times\frac{\delta^{(4)}\big(\<qi^\perp \>\eta_{ia}\big)}{\<1^\perp2^\perp\>\cdots\<n^\perp1^\perp\>}\,.
\end{equation}
All these vertices exist for any $n\geq3$.

{\em Any Coulomb-branch amplitude can be computed from the CSW-like expansion with vertices~(\ref{vert1}),~(\ref{vert2}) and~(\ref{vert3})}, connected by massive scalar propagators~(\ref{propcsw}). As in the case of two massive lines,   $q$ does not have to satisfy the constraint~(\ref{spq}) for this expansion to be valid.  However, as opposed to the two-mass case, this expansion is  somewhat less useful due to a proliferation of contributing diagrams. This, of course,  reflects the fact that Coulomb-branch amplitudes are more complicated for generic masses.

To illustrate the expansion in a simple example, consider the 4-point amplitude $\<W^-W^+W^+W^+\>$, where all four external lines are W-bosons of arbitrary mass. This amplitude is given by the sum of three diagrams:
\begin{equation}
    \big\<W_1^-W_2^+W_3^+W_4^+\big\>~~=~~
    \parbox[c]{2.3cm}{\includegraphics[width=2.3cm]{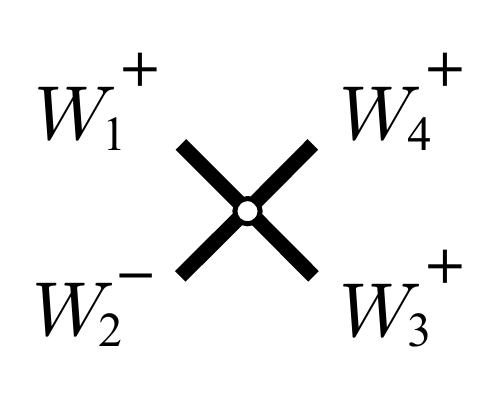}}
    ~~+~~
    \parbox[c]{2.3cm}{\includegraphics[width=2.3cm]{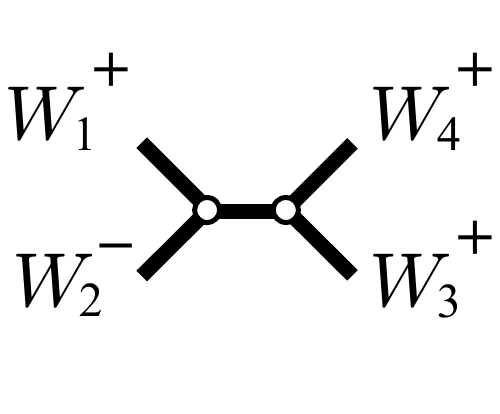}}
    ~~+~
    \parbox[c]{2.3cm}{\includegraphics[width=2.3cm]{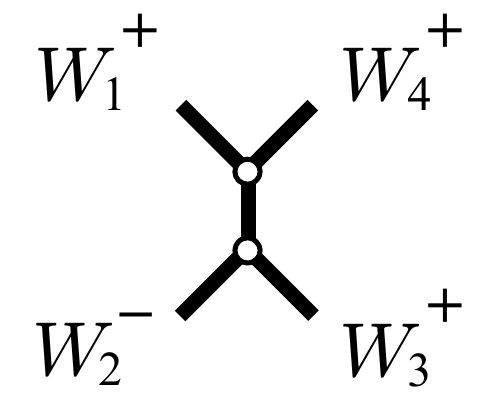}}\,.
\end{equation}
We have verified numerically that the sum of these three diagrams indeed reproduces the Coulomb-branch amplitude $\<W^-W^+W^+W^+\>$ for arbitrary values of the masses. Similarly, we have verified that the expansion reproduces the `MHV' amplitude $\<W^-W^-W^+W^+\>$ correctly for arbitrary masses. There are 5 diagrams in the expansion of this particular amplitude.

\setcounter{equation}{0}
\section{A peek at loop level}\label{secloop}
Up to now, our discussion has  strictly focused on tree amplitudes. However, much progress has also  been made in understanding the planar loop-level integrand of $\cn\!=\!4$ SYM at the origin of moduli space~\cite{Bern:1997nh,Bern:2005iz,Bern:2006ew,Bern:2007ct,ArkaniHamed:2010kv,Bullimore:2010dz,Mason:2010yk}. It is thus natural to ask whether the massless integrand also fully determines the integrand on the Coulomb branch in the planar limit.

Let us first ask which Coulomb-branch amplitudes we can in principle hope to obtain from soft-scalar limits of the {\em planar} integrand at the origin of moduli space. A diagram of the planar $L$-loop Coulomb-branch integrand will have a certain unbroken subgroup $U(M_k)\!\subset\! U(N)$ associated with each loop; this is obvious in double-line notation, as depicted in Figure~\ref{figDL}(b). We identify the gauge groups associated with the closed loop lines by  $k_1^l,k_2^l,\ldots,k_L^l$. A planar Coulomb-branch diagram loses its planarity if the soft-limit construction forces us to attach  vev scalars anywhere inside the $L$ loops; this is the case if any of the scalars in the gauge groups of loop lines have non-vanishing vevs, \ie if $v_{k_i^l}\neq 0$ for any $i$. To avoid this, we break the gauge as $U(N)\to U(M_0)\otimes \prod_k U(M_k)$, and use overall shift-invariance in the vevs to set $v_0=0$. We then take the large-$N$ limit such that  $M_0\gg M_k$. The leading contribution in this large-$N$ limit are then planar diagrams with internal loop lines in the group $U(M_0)$; we thus have $k_1^l=..=k_L^l=0$. In the following we will focus on this case.\footnote{This large-$N$ limit is closely related to the one studied in~\cite{Alday:2009zm, Henn:2010bk,Henn:2010ir,Henn:2011xk} (see also~\cite{Alday:2007hr}).}

The proposal~(\ref{generalization2}) can then be promoted to a relation between the massive and massless planar $L$-loop integrands ${\cal I}_L$ in the large-$N$ limit:
\begin{equation}\label{generalization3}
    {\cal I}_L\Bigl(X_1X_2\,\ldots\,X_n\Bigr)~~=~~\lim_{\varepsilon\to 0}\,\sum_{s=0}^\infty~\sum_{s_1+\ldots+s_n=s}\,{\cal I}_L\Bigl(\,Y_1\,
    \underbrace{\phi^{\rm vev}_{\varepsilon q}\ldotss\phi^{\rm vev}_{\varepsilon q}}_{s_1\text{ times}}\,Y_2\,
    \ldots\,Y_n
        \underbrace{\phi^{\rm vev}_{\varepsilon q}\ldotss\phi^{\rm vev}_{\varepsilon q}}_{s_n\text{ times}}\,
    \,\Bigr)_{\rm sym}\,.
\end{equation}
As in the tree-level case, the particles $X_i$ in the Coulomb-branch amplitude have mass $m_{i}=v_{k_i}\!-\!v_{k_{i-1}}$, the $Y_i$ are the corresponding massless particles as given in Table~\ref{tabstates}, and the $s_i$ vev scalars between  lines $Y_i$ and $Y_{i+1}$ are given by $\phi^{\rm vev}=v_{k_i}(\phi^{12}\!+\!\phi^{34})$.

We now generalize the  derivation of collinear and soft finiteness to the loop-integrand proposal~(\ref{generalization3}). Just like tree amplitudes, the loop integrand of massless $\cn\!=\!4$ admits an MHV vertex expansion~\cite{Brandhuber:2004yw,Bullimore:2010dz}. Most of our above analysis then immediately carries over to the loop integrand, as it did not
rely on a tree topology of the MHV vertex diagrams.  There is one subtlety that we need to address, however. Consider a diagram in the massless integrand which contains a loop that connects an MHV vertex to itself, with only vev scalars attaching to the loop; then the two lines on the MHV vertex corresponding to this loop have momenta $P_I$ and $P_J$ that only differ by a multiple of the vev scalar momentum: $P_I\!+\!P_J\sim\varepsilon q$. Naively, this leads to a new type of soft divergence from the factor $1/\<P_IP_J\>$ in the cyclic MHV denominator. However, this soft divergence is more than canceled when we perform the supersum in the numerator. Indeed, denoting $|\delta P\>=|P_I\>+|P_J\>$, we have
\begin{equation}
\begin{split}\label{cswelloop}
    \sum_{\rm states}
    \parbox[c]{1.8cm}{\includegraphics[width=1.6cm]{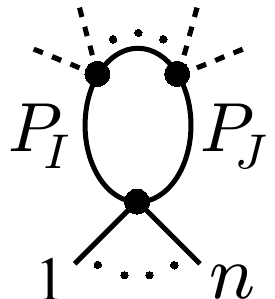}}
    ~&\propto~\int d^4\eta_{Pa}\,\frac{\delta^{(8)}\big(|P_I\>\eta_{Pa}\!+\!|P_J\>\eta_{Pa}\!+\! \sum_i|i\>\eta_{ia}\big)}{\<nP_J\>\<P_JP_I\>\<P_I 1\>\cdots\<n\dash1,n\>}
    ~=~\frac{\delta^{(4)}\big(\sum_i\<i\,\delta P\>\eta_{ia}\big)}{\<nP_J\>\<P_J\,\delta P\>\<P_I 1\>\cdots\<n\dash1,n\>}\\
    ~&=~O\big(|\delta P\>^3\big)  \,.
\end{split}
\end{equation}
This cancellation is of course well known from the massless case, where it is used to argue for the absence of MHV vertex diagrams in which a loop line starts and ends on the same vertex~\cite{Brandhuber:2005kd}. Here, in conjunction with our tree-level analysis above, it establishes the absence of soft divergences on the right-hand side of~(\ref{generalization3}).

The   proposal~(\ref{generalization3}) thus gives a finite, unambiguous prediction for the Coulomb-branch loop-integrand. As in the tree-level case, it implies the CSW-like expansion with the vertices given in section~\ref{CSW2} for two massive external lines, and in section~\ref{CSWgen} for the general case. To test the proposal, one needs to compute actual Coulomb-branch loop integrands from these rules. This is beyond the scope of the current paper.

\setcounter{equation}{0}
\section{Discussion}\label{secdisc}
In this paper we argued that  on-shell tree amplitudes on the Coulomb-branch of $\cn\!=\!4$ SYM can be computed from massless on-shell amplitudes at the origin of moduli space. We verified this proposal to all orders in the mass for a variety of amplitudes. We also proposed that an equivalent relation holds between the planar integrand of the massless theory and the Coulomb-branch integrand in a certain large-$N$ limit. As a byproduct, the analysis naturally led to a CSW-like expansion for Coulomb-branch amplitudes. 

We found it convenient in this work to use the MHV vertex expansion as a tool to show that the required massless soft-scalar amplitudes are free of divergences, and to verify the proposal for specific amplitudes; the connection between massless and massive amplitudes that we presented, however, does not rely in any way on the specific representation of the amplitude or integrand. In fact, it would be very valuable to carry out a similar analysis for other representations of the massless tree amplitudes and loop integrands; this may lead to new expressions for Coulomb-branch amplitudes  that make more symmetries manifest (such as the dual-conformal symmetry they inherit from the higher-dimensional theory~\cite{Bern:2010qa,Brandhuber:2010mm,CaronHuot:2010rj,Dennen:2010dh}).

In 6 dimensions, the relation between massless and massive amplitudes that we presented here gets unified into a relation between amplitudes in the same theory, namely massless $\cn = (1,1)$ SYM.\footnote{This theory, and its relation to $\cn\!=\!4$ SYM amplitudes in 4d, has been under intense recent investigation~\cite{Bern:2010qa,Brandhuber:2010mm,Dennen:2010dh,Hatsuda:2008pm,Elvang:2011fx,Huang:2011um}.} Examining this relation in 6d should provide a complementary and potentially more illuminating perspective on the underlying physics, in particular on the role of the $q$ constraint~(\ref{spq}). Similarly, it would also be valuable to re-derive the CSW-like expansion that arose naturally in our analysis from more conventional methods, such as the Lagrangian approach~\cite{Gorsky:2005sf,Mansfield:2005yd,Mason:2005kn}, or all-line (super)shift recursion relations~\cite{Elvang:2008vz,Kiermaier:2009yu,Cohen:2010mi,Bullimore:2010dz}. 

The methods presented in this paper are more general than the specific theory and symmetry-breaking pattern we examined here. In particular, {\em supersymmetry played no role} in our tree-level analysis. However, we relied heavily on the fact that the spontaneously-broken theory is  connected to the unbroken one through a moduli space of vacua. It is not obvious how much of our analysis carries over to theories where this is not the case. For example, it would be interesting to examine whether a CSW-like expansion is valid for amplitudes in massive QCD, possibly by generalizing the masssless analysis of~\cite{Dixon:2010ik}.

At loop level, the CSW-like expansion presented here should not only be useful for massive amplitudes, but also to compute rational terms in non-supersymmetric massless theories. Rational terms are not cut-constructible in 4 dimensions and thus elude the massless CSW expansion of loop integrands~\cite{Ettle:2007qc,Brandhuber:2007vm}. However, masses can be used to encode the $D\!>\!4$ components of loop momenta in $D$-dimensional unitarity cuts, which then recover the full loop amplitude including rational terms~\cite{Badger:2008cm}.\footnote{For massive scalars a CSW expansion was worked out in~\cite{Boels:2007pj}, which can be used to compute rational terms in QCD~\cite{Boels:2008ef,NigelGlover:2008ur}.} One technical difficulty that must be addressed, however, is the appearance of soft divergences in loop diagrams of the type displayed in~(\ref{cswelloop}), which only cancel straight-forwardly in supersymmetric theories.
A closer analysis of our proposal at loop level --- with maximal, reduced, or no supersymmetry --- seems a promising avenue for future progress.

\section*{Acknowledgments}
We thank H.~Elvang
for valuable discussions and collaboration in the initial stages of this project. We also thank N.~Beisert, Z.~Bern, L.~Dixon, D.~Freedman, D.~Skinner, T.~Slatyer and S.~Weinzierl for helpful discussions.
This research was supported in part by NSF grant PHY-0756966 (Princeton) and in part by NSF grant PHY-0551164 (KITP).

\appendix
\setcounter{equation}{0}
\section{Cancelation of soft divergences}
\subsection{Derivation of~(\ref{222})}\label{appA1}
In this appendix, we derive the  expression~(\ref{222}) for the massless amplitude ${\cal A}^{(v)}\equiv\<g^-\phi^{\rm vev}_{\varepsilon q}\ldotss\phi^{\rm vev}_{\varepsilon q}g^+\phi^{34}\phi^{34}\>_{\rm sym}$ with $s=2v+2$ vev scalars.
A  crucial ingredient in our derivation is the identity~(\ref{id}), which we repeat here for the reader's convenience:
\begin{equation}\label{idagain}
    \frac{1}{\varepsilon^t}\sum_{r=0}^t \frac{(-1)^r}{r!(t-r)!}\prod_{i=1}^N\frac{1}{1+  \varepsilon r x_i}
    ~=~ \!\!\!\!\!\sum_{t_1+\ldots+t_N=t}\,\,\prod_{i=1}^N x_i^{t_i}~+~O(\varepsilon)\,.
\end{equation}
The diagrams contributing to ${\cal A}^{(v)}$ were presented in~(\ref{sumvis}). Let us rearrange the sums in~(\ref{sumvis}) slightly, as
\begin{equation}
\begin{split}\label{rewritten}
    {\cal A}^{(v)}~=~\sum_{t=0}^v\,\sum_{r=0}^t\,~~&
       \parbox[c]{6.6cm}{\includegraphics[width=6.6cm]{subl4pt}}\,.
    \\[-6ex]
    &\hskip.5cm
    \underbrace{\hskip1.15cm}_{(t-r)\text{ times}}\hskip.43cm
    \underbrace{\hskip1.15cm}_{(v-t)\text{ times}}\hskip.41cm
    \underbrace{\hskip1.15cm}_{r\text{ times}}
\end{split}
\end{equation}
For fixed $v$, the product of MHV vertices in these diagrams is independent of $t$ and $r$ (thanks to our choice of reference spinor $|q]$ in the MHV vertex expansion):
\begin{equation}
    \prod A_{\rm MHV}~=~-(-)^{v}m^{2v+2}\frac{\<1^\perp|q|2^\perp]}{\<2^\perp|q|1^\perp]}\,.
\end{equation}
We can thus pull this factor out of the sum, and it remains to compute the sum over products of propagators.

Before we evaluate the sums over products of propagators in our specific case, it useful to stay a bit more general and compute the corresponding sum with $N$ arbitrary internal propagators, each carrying an internal momentum of the form $P_i=p_2^\perp+p_3+\ldots$\,. We find
\begin{equation}
\begin{split}\label{idprop}
  &\sum_{r=0}^t
  \Biggl[\,\prod_{i=1}^{t-r}\frac{1}{(p_1^\perp +2i\varepsilon q)^2}\Biggr]
  \Biggl[\,\prod_{i=0}^{N}\frac{1}{(P_{i}+2r\varepsilon q)^2}\Biggr]
  \Biggl[\,\prod_{i=1}^r\frac{1}{(p_2^\perp +2i\varepsilon q)^2}\Biggr]\\[1ex]
  &~=~
  \frac{1}{\varepsilon^t}\times\frac{1}{(4\,q\!\cdot\!p_1)^t}\,\sum_{r=0}^t\,\frac{(-1)^r}{r!(t-r)!}
  \,\prod_{i=1}^{N}\frac{1}{P_{i}^2\big(1+r\varepsilon \frac{4\,q\!\cdot\!P_i}{P_i^2}\big)}
  ~=~\frac{1}{(q\!\cdot\!p_1)^t}\,\sum_{t_1+\ldots+t_N=t}\,\,\prod_{i=1}^N \frac{1}{P_i^2}\Bigl(\frac{q\!\cdot\!P_i}{P_i^2}\Bigr)^{t_i}~+~O(\varepsilon)\,,
\end{split}
\end{equation}
where we used the special-$q$ constraint $q\!\cdot\!p_2=-q\!\cdot\!p_1$ in the first step, and the identity~(\ref{idagain}) in the last step.

For the case at hand, we see that the diagram in~(\ref{rewritten}) has $N\!=\!v\!-\!t\!+\!1$ propagators, with $P_i=P_{23}^\perp+2(i\!-\!1)\varepsilon q$. We can use~(\ref{idprop}) to rewrite the sum over $r$, and find
\begin{equation}
\begin{split}\label{step}
    {\cal A}^{(v)}
    ~&=~\Bigl(\prod A_{\rm MHV}\Bigr)\,\sum_{t=0}^v\,
    \frac{1}{(q\!\cdot\!p_1)^t}\,\sum_{t_1+\ldots+t_N=t}\,\,\prod_{i=1}^{v-t+1}\frac{1}{P_i^2} \Bigl(\frac{q\!\cdot\!P_i}{P_i^2}\Bigr)^{t_i}+O(\varepsilon)\\
    ~&=~\Bigl(\prod A_{\rm MHV}\Bigr)\,\frac{1}{(P_{23}^\perp)^2}\,
    \sum_{t=0}^v\,\frac{v!}{t!(v-t)!}\,\Bigl(\frac{1}{(P_{23}^\perp)^2}\Bigr)^{v-t}\Bigl(\frac{q\!\cdot\!P_{23}^\perp}{(P_{23}^\perp)^2\,q\!\cdot\! p_1}\Bigr)^{t}\, +O(\varepsilon)\,.
\end{split}
\end{equation}
Here, we used $P_i=P_{23}^\perp+O(\varepsilon)$ and
\begin{equation}
    \sum_{t_1+\ldots+t_N=t}1~~=~\frac{(t+N-1)!}{t!(N-1)!}\,.
\end{equation}
Carrying out the binomial sum in~(\ref{step}), we then find
\begin{equation}
   {\cal A}^{(v)}~=~-\frac{m^{2}\<1^\perp|q|2^\perp]}{\<2^\perp|q|1^\perp](P_{23}^\perp)^2}\biggl(\frac{m^2\,q\!\cdot\! p_3}
   {(P_{23}^\perp)^2\,q\!\cdot\! p_2}\biggr)^v\,.
\end{equation}
This is precisely the result~(\ref{222}) that we set out to derive.
\subsection{Derivation of~(\ref{vlvrsum})}\label{appA2}
In this appendix, we derive~(\ref{vlvrsum}). It is convenient to relabel the sum in~(\ref{vlvrsum}) and express it as
\begin{equation}
    \begin{split}\label{vlvrsumagain}
    \sum_{t=0}^\infty\,\sum_{r=0}^t~~
       &\parbox[c]{6.2cm}{\includegraphics[width=6.2cm]{propperpnpt}}
       ~\,~=~~\,\parbox[c]{4.1cm}{\includegraphics[width=4.1cm]{propnpt}}~+~O(\varepsilon)\,.
    \\[-5.75ex]
    &\hskip.26cm
    \underbrace{\hskip1.cm}_{(t-r)\text{ times}}\hskip2.78cm
    \underbrace{\hskip1.cm}_{r\text{ times}}
\end{split}
\end{equation}
The derivation of~(\ref{vlvrsumagain}) is straight-forward using the identity~(\ref{idprop}). In~(\ref{vlvrsumagain}), we consider a diagram with fixed internal structure, and only change the number of `dangerous' 4-point vertices $(t\!-\!r)$ and $r$ that attach to its external legs $1$ and $2$, respectively. Let us denote the $N$ (finite) internal propagators by $P_i$. The product of internal MHV vertices appears as an overall factor on both sides of~(\ref{vlvrsumagain}); in particular, this product is independent of $t$ and $r$ on the left-hand side, due to our choice of reference spinor $|q]$. This factor can thus be ignored in our analysis. Focusing on the propagators and `dangerous' four-point vertices on the left-hand side, we can use~(\ref{idprop}) to obtain
\begin{equation}
\begin{split}\label{analogous}
  &\sum_{t=0}^\infty(-m^2)^t
  \lim_{\varepsilon\to0}\sum_{r=0}^t
  \Biggl[\,\prod_{i=1}^{t-r}\frac{1}{(p_1^\perp +2i\varepsilon q)^2}\Biggr]
  \Biggl[\,\prod_{i=0}^{N}\frac{1}{(P_{i}+2r\varepsilon q)^2}\Biggr]
  \Biggl[\,\prod_{i=1}^r\frac{1}{(p_2^\perp +2i\varepsilon q)^2}\Biggr]\\[1ex]
&~=~\sum_{t=0}^\infty\,\sum_{t_1+\ldots+t_N=t}\,\,\prod_{i=1}^N \frac{1}{P_i^2}\Bigl(\,\frac{m^2\,q\!\cdot\! P_i}{P_i^2\,q\!\cdot\!p_2}\Bigr)^{t_i}
~=~\prod_{i=1}^N \frac{1}{(P_i-\frac{m^2}{2q\cdot p_2}q)^2}\,.
\end{split}
\end{equation}
We see that the summation of all divergent diagrams effectively shifts all finite internal propagators by $-m^2 q/(2 q\cdot p_2)$; this is precisely the shift $P_I^\perp\to P_I$ displayed in~(\ref{vlvrsumagain}), and thus proves the claim.

\subsection{Derivation of~(\ref{perptoprop})}\label{appA3}
In this appendix, we want to show~(\ref{perptoprop}), which we repeat here for the reader's convenience:
\begin{equation}\label{perptopropagain}
\sum_{v_1,v_2,..v_n=0}^\infty\hskip-1cm\parbox[c]{5cm}{\includegraphics[width=5cm]{snblobs}}~~~~~=~~~~~
\parbox[c]{3cm}{\includegraphics[width=3cm]{propblobs}}\,.
\end{equation}
The sum over soft-divergent diagrams should thus turn all propagators $1/(P_I^\perp)^2$ into propagators $1/P_I^2$. It is convenient to introduce
\begin{equation}
    \beta_i~\equiv~\frac{-m_i^2}{4\,q\!\cdot\! p_i}\,,\qquad \beta_{ij}~\equiv~\beta_i+\beta_{i+1}+\ldots+\beta_j\,.
\end{equation}
In this notation, the special-$q$ constraint~(\ref{spq}) is simply $\beta_{1n}=0$.
As usual, the product of internal MHV vertices is an overall factor that coincides, diagram by diagram, on the left- and right-hand sides of~(\ref{perptopropagain}), so we will ignore it in the following. The momentum on the internal line $I$ is given by
\begin{equation}
    P_{I,\varepsilon}^\perp~\equiv~\sum_{i\in I}\bigl[p^\perp_i+2\varepsilon v_i q\bigr]\,.
\end{equation}
What we need to show is
\begin{equation}\label{toshow}
    {\cal A}~=~\prod_I\frac{1}{P_{I}^2}+O(\varepsilon)\,,
\end{equation}
with
\begin{equation}
    {\cal A}~\equiv~\sum_{v_1,..v_n=0}^\infty\frac{\beta_1^{v_1}\cdots \beta_n^{v_n}}{\varepsilon^{v_1+..+v_n}v_1!\cdots v_n!}
    \prod_I\frac{1}{( P_{I,\varepsilon}^\perp)^2}\,.
\end{equation}
It is convenient to make the following change of variables in the sum:
\begin{equation}
\begin{split}
    {\cal A}
    ~&=~\sum_{v_1,..v_n=0}^\infty\frac{\beta_1^{v_1}(\beta_{12}-\beta_1)^{v_2}(\beta_{13}-\beta_{12})^{v_3}\cdots (\beta_{1n}-\beta_{1,n-1})^{v_n}}{v_1!\cdots v_n!\,\varepsilon^{v_1+..+v_n}}
    \prod_I\frac{1}{( P_{I,\varepsilon}^\perp)^2}\\
    ~&=~\!\!\!\!\!\!\!\sum_{v_1,v_1',v_{12},..v_{1,n\dash1}'=0}^\infty\!\!\!\!\!\frac{(-)^{v_1'}\beta_1^{v_1+v_1'}(-)^{v_{12}'}\beta_{12}^{v_{12}+v_{12}'}(-)^{v_{13}'}\beta_{13}^{v_{13}+v_{13}'}\cdots (-)^{v_{1,n\dash1}'}\beta_{1,n\dash1}^{v_{1,n\dash1}+v_{1,n\dash1}'}}{v_1!v_1'!v_{12}!\cdots v_{1,n\dash1}'!\,\varepsilon^{v_1+v_1'+v_{12}+..+v_{1,n\dash1}'}}
    \prod_I\frac{1}{( P_{I,\varepsilon}^\perp)^2}\,,
\end{split}
\end{equation}
where we used $\beta_{1n}=0$ and expanded
\begin{equation}
    \frac{(\beta_{1i}-\beta_{1,i\dash1})^{v_i}}{v_i!}~=\sum_{v_{1,i\dash1}'+v_{1i}=v_i}\!\!\!\frac{(-\beta_{1,i\dash1})^{v_{1,i\dash1}'}\beta_{1i}^{v_{1i}}}{v_{1,i\dash1}'!v_{1i}!}\,.
\end{equation}
To make contact with~(\ref{idprop}), we make another change of variables in the sum by defining $t_i=v_{1i}+v_{1i}'$ and $r_i=v_{1i}'$. We then have
\begin{equation}
\begin{split}\label{ready}
      {\cal A}
    ~&=~\sum_{t_1,t_2,..t_{n\dash1}'=0}^\infty\,\,\sum_{r_1=0}^{t_1}\cdots\!\!\!\sum_{r_{n\dash1}=0}^{t_{n\dash1}}
    \frac{(-)^{r_1}\beta_1^{t_1}}{r_1!(t_1-r_1)!\varepsilon^{t_1}}\cdots \frac{(-)^{r_{n\dash1}}\beta_{1,n\dash1}^{t_{n\dash1}}}{
    r_{n\dash1}!(t_{n\dash1}-r_{n\dash1})!\,\varepsilon^{t_{n\dash1}}}
    \prod_I\frac{1}{( P_{I,\varepsilon}^\perp)^2}\,.
\end{split}
\end{equation}
We now express the momenta $P_{I,\varepsilon}^\perp$ in terms of the new summation variables:
\begin{equation}\label{newPIeps}
    P_{I,\varepsilon}^\perp
    ~=~\sum_{i\in I}\bigl[p^\perp_i+2\varepsilon (t_i-r_i+r_{i\dash 1}) q\bigr]\,,\qquad\text{with }r_0\equiv r_n\equiv t_n\equiv0\,.
\end{equation}
Let us now  carry out the sums in~(\ref{ready}) over  $r_1$ and $t_1$ using~(\ref{idprop});
it is easy to see that, to leading order in $\varepsilon$, this amounts to setting $t_1=0$ in all $P_{I,\varepsilon}^\perp$, and replacing $\varepsilon r_1 \to -\beta_1$. In fact, we can literally follow the steps of the computation~(\ref{analogous}) in the two-mass case, where~(\ref{idprop}) was used to show that the sum over $r$ and $t$  amounts to the replacement $\varepsilon r\to -\beta$, with $\beta=-\frac{m^2}{2 q\cdot p_2}$. Proceeding iteratively, the sum over $r_i$ and $t_i$ amounts to setting $t_i=0$ in all $P_{I,\varepsilon}^\perp$, and replacing $\varepsilon r_i\to -\beta_{1i}$. After all sums, we have thus replaced
\begin{equation}
     P_{I,\varepsilon}^\perp~\to~\sum_{i\in I}\bigl[p^\perp_i+2(\beta_{1i}-\beta_{1,i\dash 1}) q\bigr]
     ~=~\sum_{i\in I}\Bigl[p^\perp_i-\frac{m_i^2}{2\,q\!\cdot\! p_i} q\Bigr]~=~P_I\,.
\end{equation}
This proves~(\ref{toshow}) and thus completes our derivation.

\small
\renewcommand{\baselinestretch}{.9}

%\bibliographystyle{utphys}
%\bibliography{coulombbib}{}

\providecommand{\href}[2]{#2}\begingroup\raggedright\endgroup

\end{document}